\documentclass[12pt]{article}
\usepackage{epsfig,amsmath,amsfonts,amssymb,makeidx,ifthen}
\newtheorem{theorem}{Theorem}[section]
\newtheorem{lemma}[theorem]{Lemma}

\newtheorem{proposition}[theorem]{Proposition}
\makeatletter
\@addtoreset{equation}{section}
\makeatother

\makeatletter
\long\def\@makecaption#1#2{{\small
\advance\leftskip1cm
\advance\rightskip1cm
\vskip\abovecaptionskip
\sbox\@tempboxa{#1: #2}%
\ifdim \wd\@tempboxa >\hsize
 #1: #2\par
\else
\global \@minipagefalse
\hb@xt@\hsize{\hfil\box\@tempboxa\hfil}%
\fi
\vskip\belowcaptionskip}}
\makeatother
\def\eq#1\en{\begin{equation}#1\end{equation}}  
\def\eqa#1\ena{\begin{align}#1\end{align}}
\def\eqg#1\eng{\begin{gather}#1\end{gather}}
\newcommand{\lb}[1]{\label{e:#1}}
\newcommand{\rlb}[1]{\eqref{e:#1}} 
\newcommand{\nl}{\notag\\}
\newcommand{\nnb}	{\nonumber \\} 
\newcommand{\itext}{\notag\intertext}
\def\eqalign#1\enalign{
	\begin{align}#1\end{align}
	}
\newcommand{\sbkt}[1]{\langle#1\rangle}
\newcommand{\bbkt}[1]{\bigl\langle#1\bigr\rangle}

\newcommand{\sumtwo}[2]%
{\mathop{\sum_{#1}}_{#2}}
\newcommand{\sumthree}[3]%
{\mathop{\mathop{\sum_{#1}}_{#2}}_{#3}}
\newcommand{\sumfour}[4]%
{\mathop{\mathop{\mathop{\sum_{#1}}_{#2}}_{#3}}_{#4}} 
\newcommand{\prodtwo}[2]%
{\mathop{\prod_{#1}}_{#2}}
\newcommand{\mintwo}[2]%
{\mathop{\min_{#1}}_{#2}}
\newcommand{\maxtwo}[2]%
{\mathop{\max_{#1}}_{#2}}
\newcommand{\maxthree}[3]%
{\mathop{\mathop{\max_{#1}}_{#2}}_{#3}}
\newcommand{\limtwo}[2]%
{\mathop{\lim_{#1}}_{#2}}
\newcommand{\suptwo}[2]%
{\mathop{\sup_{#1}}_{#2}}
\newcommand{\supthree}[3]%
{\mathop{\mathop{\sup_{#1}}_{#2}}_{#3}}
\newcommand{\supfour}[4]%
{\mathop{\mathop{\mathop{\sup_{#1}}_{#2}}_{#3}}_{#4}} 
\newcommand{\inftwo}[2]%
{\mathop{\inf_{#1}}_{#2}}
\newcommand{\infthree}[3]%
{\mathop{\mathop{\inf_{#1}}_{#2}}_{#3}}
\newcommand{\inffour}[4]%
{\mathop{\mathop{\mathop{\inf_{#1}}_{#2}}_{#3}}_{#4}} 
\newcommand\calB{{\cal B}}

\newcommand\calH{{\cal H}}

\newcommand\calN{{\cal N}}

\newcommand\calU{{\cal U}}

\newcommand{\bspsi}{\boldsymbol{\psi}}
\newcommand{\bsphi}{\boldsymbol{\varphi}}
\newcommand{\bsxi}{\boldsymbol{\xi}}

\newcommand{\sfA}{\mathsf{A}}
\newcommand{\sfB}{\mathsf{B}}

\newcommand{\sfP}{\mathsf{P}}
\newcommand{\sfQ}{\mathsf{Q}}

\newcommand{\sfS}{\mathsf{S}}

\newcommand{\sfU}{\mathsf{U}}

\newcommand{\bbR}{\mathbb{R}}

\newcommand{\up}{\uparrow}

\newcommand{\qedm}{\hfill\mbox{$\blacksquare$}}
\newcommand{\Hneq}{\calH_\mathrm{neq}}
\newcommand{\Heq}{\calH_\mathrm{eq}}
\newcommand{\Hrnd}{\calH_\mathrm{rnd}}
\newcommand{\Pneq}{\hat{P}_\mathrm{neq}}
\newcommand{\Prnd}{\hat{P}_\mathrm{rnd}}
\newcommand{\DU}{\mathit{\Delta}U}
\newcommand{\DE}{\mathit{\Delta}E}
\newcommand{\hH}{\hat{H}}

\newcommand{\hU}{\hat{U}}
\newcommand{\tE}{\tilde{E}}
\newcommand{\tauB}{\tau_\mathrm{B}}
\newcommand{\taum}{\tau_\mathrm{max}}
\newcommand{\tbeta}{\tilde{\beta}}

\newcommand{\oD}{\{1,\ldots,D\}}
\newcommand{\ab}{{\alpha\beta}}
\newcommand{\suma}{\sum_{\alpha=1}^D}
\newcommand{\sumab}{\sum_{\alpha,\beta=1}^D}
\newcommand{\psa}{\bspsi_\alpha}
\newcommand{\psb}{\bspsi_\beta}

\newcommand{\prob}{\mathbb{P}}
\newcommand{\Av}{\mathbb{E}}
\newcommand{\tr}{\mathrm{Tr}}
\newcommand{\Wg}{{\mathrm{Wg}}}

\newcommand{\Sn}{{\mathfrak{S}_n}}
\newcommand{\idp}{\mathrm{id}}
\newcommand{\shif}{\sigma_1}

\newcommand{\mumax}{\mu_\mathrm{max}}

\newcommand{\bmu}{\bar{\mu}}
\newcommand{\lamax}{\lambda_\mathrm{max}}

\newcommand{\alphan}{\nu}
\newcommand{\betan}{\zeta}
\newcommand{\Eab}{E_\alpha-E_\beta}
\newcommand{\taue}{\tau_\mathrm{esc}}

\newcommand{\bigno}{\par\bigskip\noindent}
\newcommand{\newremark}{\par\vspace{4pt}\noindent}

\begin{document}

\noindent
{\bf\large 
The approach to equilibrium in a macroscopic quantum system for a typical nonequilibrium subspace
}
\par\bigskip

{\small
\noindent 
Sheldon Goldstein\footnote{
Departments of Mathematics and Physics, Rutgers University, 110 Frelinghuysen Road, Piscataway, NJ 08854-8019, USA
}, 
Takashi Hara\footnote{
Faculty of Mathematics,
Kyushu University,
Moto-oka, Nishi-ku,
Fukuoka 819-0395,
Japan
},
and Hal Tasaki\footnote{
Department of Physics, Gakushuin University, Mejiro, Toshima-ku, 
Tokyo 171-8588, Japan
}
}

%%%%%%%%%%%%%%%%%%%%%%%%%%%%%%%%%%%%%%
%%%%%%%%%%%%%%%%%%%%%%%%%%%%%%%%%%%%%%

%%%%%%%%%%%%%%%%%%%%%%%%%%%%%%%%%%%%%%
\begin{abstract}
We study the problem of the approach to equilibrium in a macroscopic quantum system in an abstract setting.
We   prove that, for a typical choice of ``nonequilibrium subspace'', any initial state (from the energy shell)  thermalizes, and in fact does so very quickly, on the order of the Boltzmann time $\tauB:=h/(k_\mathrm{B}T)$.
This apparently unrealistic, but mathematically
rigorous, conclusion has the important physical implication that the moderately slow decay
observed in reality is not typical in the present setting.

The fact that macroscopic systems approach thermal equilibrium may seem puzzling, for example, because it may  seem to conflict with the time-reversibility of the microscopic dynamics.
According the present result, what needs to be explained is, not that macroscopic systems approach equilibrium, but that they do so slowly.

Mathematically our result is based on an interesting property of the maximum eigenvalue of the Hadamard product of a positive semi-definite matrix and a random projection matrix.
The recent exact formula by Collins for the integral with respect to the Haar measure of the unitary group plays an essential role in our proof.
\end{abstract}

\tableofcontents

%%%%%%%%%%%%%%%%%%%%%%%%%%%%%%%%%%%%%%%%%%%
%%%%%%%%%%%%%%%%%%%%%%%%%%%%%%%%%%%%%%%%%%%
\section{Introduction}
\label{s:intro}
Recently there has been a considerable renewed interest in the foundation of quantum statistical mechanics.
This has led, in particular, to a revival of the old approach by von Neumann to investigate the problem of thermalization only in terms of quantum dynamics in an isolated system \cite{vonNeumann,GLMTZ10,GLTZ}.
It has been demonstrated in some general or concrete settings that a pure initial state evolving under quantum dynamics indeed thermalizes (i.e., approaches thermal equilibrium)  in a certain mathematical  sense\footnote{%
\label{fn:twoclasses}
These rigorous results about thermalization (or equilibration) can be divided into two classes.
In the first class, which goes back to \cite{vonNeumann,GLMTZ10,GLTZ} and includes \cite{GLMTZ09b,Hal2010}, one is allowed to take any initial state (from the energy shell).
In the second class, which (as far as we know) starts from \cite{Hal1998} and includes \cite{Reimann,LindenPopescuShortWinter,ReimannKastner,Reimann2},
one uses initial states which have sufficiently broad energy distribution.
There is an essential difference in the mechanisms of equilibration in the two classes.
See Appendix~\ref{app-thermalization}.
}$^,$\footnote{
\label{fn:thermalizationEquilibration}
Most of these works \cite{Hal1998,Reimann,LindenPopescuShortWinter,ReimannKastner,Reimann2,SatoKanamotoKaminishiDeguchi} discuss thermalization (or equilibration) by examining whether the quantum mechanical expectation values of certain observables come and stay close to the corresponding equilibrium values.
This is different from the approach in the present work, which is based on a decomposition of the Hilbert space.
} \cite{Hal1998,Reimann,LindenPopescuShortWinter,
GLMTZ09b,Hal2010,ReimannKastner,Reimann2,SatoKanamotoKaminishiDeguchi}.
The underlying related idea that a typical pure state of a macroscopic quantum system can fully describe  thermal equilibrium has also become much more concrete
\cite{PopescuShortWinter,GLTZ06,Sugita,SugiuraShimizu12,SugiuraShimizu13}.
An important issue then is to understand the time scale necessary for thermalization \cite{ShortFarrelly,VinayakZnidaric,Masanes,Brandao,Cramer,GHT13,GHT13B,Malabarbaetal,Monnai}.

Let us briefly describe the setting and the main result.
Precise definitions will be given in later sections.
We here follow the approaches of von Neumann \cite{vonNeumann,GLMTZ10,GLTZ} and of Goldstein, Lebowitz, Mastrodonato, Tumulka, and Zangh\`\i\ \cite{GLMTZ09b} (see also \cite{Hal2010,GHT13}), and study the problem of the approach to equilibrium of an isolated macroscopic quantum system\footnote{%
Of course there is no such thing in reality as a completely isolated system.
Our motivation for studying isolated systems is basically to learn what physics (including thermodynamic behavior and the approach to equilibrium) can be realized in isolated systems.
We can study the roles played by the interaction with surrounding environment after that.
} in an abstract setting.

We consider the energy shell $\calH$, which is a linear space spanned by energy eigenstates corresponding to the narrow energy range $[U-\DU,U]$, and assume that nonequilibrium states of the system are characterized by a subspace $\Hneq$ of $\calH$.
We also assume that the dimension $d$ of $\Hneq$ is much smaller than the dimension $D$ of the energy shell $\calH$.
We regard states close to $\Hneq$ as being  out of equilibrium and those far from $\Hneq$ as in equilibrium.

Then the question of the approach to equilibrium is formulated as follows.
One starts from an initial state $\bsphi(0)\in\calH$ which may be close to $\Hneq$.
One then asks how the expectation value $\sbkt{\bsphi(t),\Pneq\,\bsphi(t)}$ behaves as a function of time $t$, where $\Pneq$ is the orthogonal projection onto $\Hneq$ and $\sbkt{\cdot,\cdot}$ denotes the inner product.
If one finds that $\sbkt{\bsphi(t),\Pneq\,\bsphi(t)}\ll1$ for some $t$, it means that $\bsphi(t)$ is far from $\Hneq$ and hence in equilibrium.

In reality the subspace $\Hneq$ should be almost uniquely determined from physical properties of the system.
It is not an easy task, however, to characterize $\Hneq$ in general or in a concrete setting. 
Nor is it easy to usefully estimate the relaxation time for any specific physical $\Hneq$. 
We shall therefore take an abstract approach in which we regard $\Hneq$ as a general $d$-dimensional subspace of $\calH$, and try to elucidate the relation between $\Hneq$ and the associated relaxation time.

In our previous work \cite{GHT13}, we proved via an explicit (and artificial)
construction that, depending on the choice of nonequilibrium subspace $\Hneq$, the relaxation time can be
as extremely large as $h\,d/\DU$, exceeding the age of the universe, or as ridiculously short as $h/\DU$, where $h$ is Planck's constant.

This motivates us to study the time scale of thermalization, or, more precisely, that of the escape from the nonequilibrium subspace $\Hneq$, for various choices of $\Hneq$.
It may be natural to first focus on a setting in
which the ``nonequilibrium subspace'' $\Hneq$ is not, in fact, a realistic nonequilibrium subspace, but rather is chosen in a
completely random manner.
One might hope that for such a subspace one
 generically has a realistic relaxation time.
If this were so, it would seem reasonable to believe that the same thing would probably be true for a realistic nonequilibrium subspace (unless we have reasons to expect otherwise).
This is basically von Neumann's philosophy in \cite{vonNeumann}.
Unfortunately this hope turns out to be far too
optimistic.

In the present paper we study the setting where $\Hneq$ is drawn randomly, and prove that, with probability close to one, the expectation $\sbkt{\bsphi(t),\Pneq\,\bsphi(t)}$ quickly becomes extremely small (when averaged in time) for \emph{any} initial state $\bsphi(0) \in \calH$.
This means that any state (including those that are very far from equilibrium) quickly approaches equilibrium, provided that we interpret $\Hneq$ as the physical nonequilibrium subspace.
The time necessary for the thermalization is of order the Boltzmann time $\tauB:=h/(k_\mathrm{B}T)$, which is usually extraordinarily short.
At room temperature, for example, $\tauB$ is of order $10^{-13}\rm\,s$.
We thus have a mathematically rigorous theorem that basically tells us that our coffee is no longer hot after, say, a micro-second\footnote{%
One may remark that our coffee cup is not an isolated quantum system.
In this case, one should regard the whole room containing the cup as a single macroscopic system, and assume that it is isolated from the outside world.
}!

This conclusion is of course absurd and highly unphysical\footnote{%
It is true, however, that the Boltzmann time $\tauB$ is the characteristic time scale for various quantum phenomena.
}.
But it indeed has the deep physical implication that a realistic system (in which coffee is hot even after a few minutes) are not covered by our theorem.
In other words, we can conclude that {\em the moderately slow decay
observed in reality is not typical}\/ in the present setting, where we draw $\Hneq$ randomly.

The method of appeal to typicality in quantum physics was (probably) initiated by von Neumann in \cite{vonNeumann} and played crucial roles in many situations including successful physical applications of the theory of random matrices.
The method is clearly summarized in \cite{GLTZ} as\footnote{%
From Section 6 of \cite{GLTZ}.
We have made small modifications to make the quote consistent with the present discussion.
}
\begin{quote}
It means that, if a property is true of most $\Hneq$, this fact may suggest that the property is also true of a concrete given system, unless we have reasons to expect otherwise.
\end{quote}

We have now come up with a  nontrivial and interesting situation in which the method of typicality does not work as intended.
In our case the ``property that is true of most $\Hneq$'', i.e., a very quick decay, turns out to be simply unphysical.
Therefore, in order to deal with the problem of thermalization in an isolated quantum system, one can no longer rely on a crude appeal to typicality, but should develop new points of view and methods which take into account some essential features of realistic macroscopic systems.
We point out that recent works \cite{GHT13,Malabarbaetal} on the time-scale of thermalization (or equilibration) may contain hints for such future directions.
See section~\ref{s:physics} for further discussions.

\bigskip
From a mathematical point of view, our result is based on the following interesting  property about the maximum eigenvalue of the Hadamard product (or the Schur product) of a positive semi-definite matrix and a random projection matrix.

Let $\sfA=(A_\ab)_{\alpha,\beta\in\oD}$ be a $D \times D$ positive semi-definite matrix with $A_{\alpha\alpha}\le1$ for all $\alpha\in\oD$.
We denote by $\mumax$ the maximum eigenvalue of $\sfA$.
Let $\sfP=(P_\ab)_{\alpha,\beta\in\oD}$ be the projection matrix onto a randomly drawn $d$-dimensional subspace, where $d\ll D$.
We define a new matrix $\sfB$ as their Hadamard product, i.e., $B_\ab:=A_\ab P_\ab$, and denote by $\lamax$ the maximum eigenvalue of $\sfB$.
We then prove a bound for $\lamax$ which roughly implies that  $\lamax\lesssim\mumax/D$ with probability very close to one\footnote{%
Throughout the present paper, $A\sim B$ means that $A/B$ is close to one.
$A\approx B$ means the weaker relation that $A/B$ is $O(1)$.
Likewise, $A\lesssim B$ means $A\le B'$ with $B'\sim B$.
}.
See Proposition~\ref{t:main3cor} for the precise statement.

The recent exact formula by Collins \cite{Coll03} of the integral with respect to the Haar measure of the unitary group plays an essential role in our proof.

\bigskip
In \cite{Cramer}, Cramer proved a closely related result, which may be interpreted as an infinite temperature version of ours\footnote{%
Cramer considers a random Hamiltonian.
Our results apply automatically to the setting where
we fix the nonequilibrium subspace and draw the (eigenbasis of the) Hamiltonian randomly.
See the remark below Proposition~\ref{c:mumax} for a further relation between Cramer's work and ours.
}.
See also \cite{VinayakZnidaric,Masanes,Brandao}.

In a very recent work \cite{Malabarbaetal}, Malabarba, Garc\'{i}a-Pintos, Linden, Farrelly, and Short studied the equilibration in quantum systems, and proved, among other things, that most observables equilibrate quite rapidly.
This result is of course quite similar to ours\footnote{
There are however essential differences between their work and ours.
The initial state needs to have a sufficiently broad energy distribution in the former, while it is arbitrary in the latter.
The observable is drawn randomly {\em after}\, fixing the initial state in the former, while the random subspace is fixed at the beginning in the latter.
The former treats equilibration while the latter deals with thermalization.
}.

The main result of the present paper and the ideas of the proof were already announced in \cite{GHT13B}.

%%%%%%%%%%%%%%%%%%%%%%%%%%%%%%%%%%%%%%%%%%%
%%%%%%%%%%%%%%%%%%%%%%%%%%%%%%%%%%%%%%%%%%%
\section{Main result and its implications}

%%%%%%%%%%%%%%%%%%%%%%%%%%%%%%%%%%%%%%%%%%%
\subsection{Setting and some background}
\label{s:BG}
We consider an abstract model of an isolated macroscopic quantum system in a large volume. 
A typical example is a system of $N$ particles confined in a box of volume $V$, where the density $N/V$ is kept constant when $V$ becomes large.
In what follows we assume that the volume $V$ is fixed, and do not discuss the $V$ dependence of various quantities explicitly.

Let $\hH$ be the Hamiltonian, and denote by $E_\alpha$ and $\psa$ the eigenvalues and  the normalized eigenstates, respectively, of $\hH$, i.e.,
\eq
\hH\psa=E_\alpha\psa.
\lb{Hp=Ep}
\en

We focus on the energy interval $[U-\DU,U]$, where $\DU$ denotes a range of energy which is small from the macroscopic point of view but is still large enough to contain many energy levels.
The choice of $\DU$ is somewhat arbitrary.
It is convenient to relabel the index $\alpha$ so that the energy eigenvalues $E_\alpha\in[U-\DU, U]$  precisely correspond to the indices $\alpha\in\oD$.
We shall work with the Hilbert space $\calH$ spanned by all $\psa$ with $\alpha\in\oD$, which is often called the {\em microcanonical energy shell}\/.

\bigskip
Let us briefly describe the problem of the approach to equilibrium, mainly following \cite{GLMTZ10,GLTZ,GLMTZ09b,Hal2010}.
We recommend \cite{GLTZ} as an accessible exposition.

It has been well established by now that, in a normal macroscopic quantum system, the overwhelming majority of states in the energy shell $\calH$ correspond to the thermal equilibrium state 
\cite{GLMTZ10,GLTZ,PopescuShortWinter,GLTZ06,Sugita}.

To formulate this fact mathematically\footnote{%
We do not mean that this is the only possible formulation.
}, we assume that the energy shell $\calH$ is decomposed into the equilibrium and the nonequilibrium subspaces as $\calH=\Heq\oplus\Hneq$.
We regard any state $\bsphi$ which is close enough to $\Heq$ as being an equilibrium
state, and a state not close to $\Heq$ as a nonequilibrium state\footnote{%
Consider the simplest situation where one is interested  in the behavior of a single macroscopic quantity $\hat{O}$, whose equilibrium value is $\bar{O}$.
Then one can define $\Heq$ as the subspace spanned by the eigenstates of the nonnegative operator $(\hat{O}-\bar{O})^2$ corresponding to sufficiently small eigenvalues.
}.
Note that neither the set of equilibrium states nor that of nonequilibrium states forms a subspace of $\calH$.
We assume that the subspace $\Heq$ occupies most of the energy shell $\calH$ in the sense that the dimension $d$ of the nonequilibrium subspace $\Hneq$ satisfies $d\ll D$.
One then easily finds that a typical state in the energy shell is an  equilibrium state\footnote{
Let  $\Pneq$ be the projection onto $\Hneq$, and consider the expectation value $\sbkt{\bsphi,\Pneq\,\bsphi}$.
By taking the uniform average over all normalized  $\bsphi\in\calH$, we get $\Av[\sbkt{\bsphi,\Pneq\,\bsphi}]=d/D\ll1$.
From the standard argument based on the Markov inequality, we find that $\sbkt{\bsphi,\Pneq\,\bsphi}\ll1$ for a typical $\bsphi\in\calH$.
The inequality $\sbkt{\bsphi,\Pneq\,\bsphi}\ll1$ implies  $\bsphi$ is very close to $\Heq$ and hence is an equilibrium state.
}.

The dimensions $D$ and $d$ are usually huge, and are expected to depend on the volume $V$ as $D\approx e^{\sigma V}$ and $d\approx e^{\sigma'V}$ with constants (entropy densities) $0<\sigma'<\sigma$.
We note however that it is not easy to actually prove this property for nontrivial quantum many-body systems.
See section~I.B of \cite{GLMTZ09b}.

The next question is whether an isolated quantum system evolving under the unitary time evolution exhibits  the approach to equilibrium.
We assume that the system starts from a normalized initial state $\bsphi(0)\in\calH$ which may not be close to $\Heq$, and ask  whether its time evolution
\eq
\bsphi(t)=e^{-i\hH t/\hbar}\bsphi(0), 
\lb{phit}
\en
where $t\ge0$, comes and stays, for most $t$, very close to $\Heq$ when $t$ is large.

Let $\Pneq$ denote the orthogonal projection onto the nonequilibrium subspace $\Hneq$.
In some settings (and under suitable assumptions), it has been proved that \cite{Hal1998,Reimann,LindenPopescuShortWinter,
GLMTZ09b,Hal2010,ReimannKastner,Reimann2}
\eq
\frac{1}{\tau}\int_0^\tau dt\,\sbkt{\bsphi(t),\Pneq\,\bsphi(t)}\ll1
\lb{approach}
\en
for sufficiently large $\tau>0$.
See Appendix~\ref{app-thermalization}.
In the present paper we also prove a statement of this type.
The bound \rlb{approach} implies that, within the time interval $[0,\tau]$,  the state $\bsphi(t)$ spends most of the time in the close vicinity of the  equilibrium subspace $\Heq$, i.e., the system approaches equilibrium within the time scale of order $\tau$.
See section~\ref{s:physics}.

Let us briefly recall a version of such results proved by Goldstein, Lebowitz, Mastrodonato, Tumulka, and Zangh\`\i\ \cite{GLMTZ09b}, who followed the philosophy of von Neumann's \cite{vonNeumann,GLMTZ10,GLTZ}.
As we discussed in the introduction, we here regard the nonequilibrium subspace $\Hneq$ as a general $d$-dimensional subspace of $\calH$.
Then it was proved that, for a typical choice\footnote{%
In \cite{GLMTZ09b}, the authors fix $\Hneq$ and choose the orthonormal basis $\{\bspsi_\alpha\}_{\alpha\in\oD}$ randomly.
But this is equivalent to the present formulation.
See also \cite{Hal2010}.
} of $\Hneq$, one has
\eq
\sbkt{\bspsi_\alpha,\Pneq\,\bspsi_\alpha}\ll1
\en
for any $\alpha\in\oD$.
This is a version of the property usually called ``energy-eigenstate thermalization''.
By using this property, it was shown that \rlb{approach} is valid for {\em any}\/ initial state $\bsphi(0)\in\calH$ (see Theorem~\ref{t:app-th-1}). 

Although this seems to establish the desired approach to equilibrium, we would also like to have some information about how large $\tau$ should be in \rlb{approach}.
We have treated this problem of time scale explicitly in \cite{GHT13}, and proved two theorems by explicit (and purely mathematical) construction of $\Hneq$; for some choices of $\Hneq$ the required time scale becomes as large as $\tau\approx h\,d/\DU$ which can easily exceed the age of the universe, while for some other choices it becomes as small as $\tau\approx h/\DU$ which is ridiculously short.

This observation suggests that our understanding of the approach to equilibrium in isolated quantum systems is still quite primitive.
In particular we need to learn more about the time scale required for thermalization.

%%%%%%%%%%%%%%%%%%%%%%%%%%%%%%%%%%%%%%%%%%%
\subsection{Assumptions and main result}
\label{s:mainresult}
We shall state our main result precisely in the present subsection.

We assume that the distribution of the energy eigenvalues 
$E_1,\ldots,E_D\in[U-\DU, U]$ is well-described by a single function $\rho(E)$, 
the density of states.
More precisely, we assume\footnote{
To be precise \rlb{sumtoint} is not an assumption since it can always be   satisfied.
See Appendix~\ref{app-density}.
The real assumption, which is contained in \rlb{rEbD}, is that the function $\rho(E)$ behaves as a physical density of states, which is 
smooth and (almost) monotone increasing. 
} for any differentiable function $f(E)$ that
\eq
	\frac{1}{D}\biggl|\suma f(E_\alpha)-\int_{U-\DU-\eta}^{U+\eta} dE\,
	\rho(E)\,f(E)\biggr|
	\le \eta \times \sup_{E\in[U-\DU-\eta,U+\eta]}|f'(E)|,
\lb{sumtoint}
\en
where $\eta$ is a small constant.
See Appendix~\ref{app-density} for details. 
Here we take the constant as
\eq
	\eta = D^{-\kappa}\tbeta^{-1}.
	\lb{eta-def}
\en  
where $\kappa\in(0,1)$ is a constant close to 1.

We shall make an essential assumption that $\rho(E)$ satisfies 
\eq
	\rho(E)\le\tbeta D
\lb{rEbD}
\en
for any $E\in[U-\DU-\eta,U+\eta]$. 
The motivation for this upper bound is explained in Remark 2 at the end of the section.
The constant $\tbeta$ which appears both in \rlb{eta-def} 
and \rlb{rEbD} is interpreted as the inverse temperature 
$\tbeta=(k_\mathrm{B}T)^{-1}$ corresponding to the 
equilibrium state in the energy shell.
Here $k_\mathrm{B}\sim1.38\times10^{-23}\rm\,J/K$ is the Boltzmann constant and $T$ is the absolute temperature.

We also define the {\em Boltzmann time}
\eq
\tauB:=h\tbeta=\frac{h}{k_\mathrm{B}T},
\lb{tauB}
\en
which is a natural time scale associated with the absolute temperature $T$.
Here $h=2\pi\hbar\sim6.626\times10^{-34}\rm\,Js$ is the Planck constant.
Note that the Boltzmann time is extremely short for practical temperatures.
For example $\tauB\approx1.6\times10^{-13}\rm\,s$ for the room temperature $T\approx300\rm\,K$, and $\tauB\approx0.5\rm\,s$ for $T\approx10^{-10}\rm\,K$, which is the lowest possible temperature that can be achieved in the current ultracold atom experiments.

In order to investigate the properties of a generic ``nonequilibrium subspace'', we shall generate a $d$-dimensional subspace of the energy shell $\calH$ randomly as follows.
In what follows, we denote the random subspace as $\Hrnd$ and the corresponding projection as $\Prnd$ to emphasize that these are random objects.
Note that $\Hrnd$ is our probabilistic model of $\Hneq$.

By $\calU(\calH)$ we denote the group of all the
unitary transformations on $\calH$.
For each $\hU\in\calU(\calH)$ and $j\in\{1,\ldots,d\}$, we define the normalized state $\bsxi^{(j)}=\hU\bspsi_j$.
We then consider the $d$-dimensional subspace $\Hrnd$ spanned by $\bsxi^{(1)},\ldots,\bsxi^{(d)}$, and the corresponding projection operator 
\eq
\Prnd=\sum_{j=1}^d\hat{p}[\bsxi^{(j)}], 
\lb{Pdef}
\en
where $\hat{p}[\bsxi]$ denotes the orthogonal projection onto $\bsxi$.
By drawing $\hU\in\calU(\calH)$ according to the unique Haar measure on $\calU(\calH)$, we can generate the $d$-dimensional subspace $\Hrnd$ and the associated projection $\Prnd$ in a completely random manner.

Our main finding is summarized in the following theorem.
The theorem is proved in section~\ref{s:pre} based on Proposition~\ref{t:main3cor}, which is valid for more general matrices. 
%%%%% THM %%%%%
\begin{theorem}
\label{t:main}
Fix an arbitrary (small) $\varepsilon>0$.
Let the dimension $D$ be sufficiently large and the dimension $d\ge1$ be sufficiently small compared with $D$, so that
\eq
	\frac{d}{D} \leq 2 \biggl (
	\frac{\log(2 e^3 D^{5/4})}{\log(1+\varepsilon)}+1 
	\biggr )^{-4} .
\lb{main-cond-on-d_and_D}
\en 
Let $\Hrnd$ and $\Prnd$ be the random $d$-dimensional subspace and the corresponding projection as defined above.
Then, with probability larger than $1-(d/D)$, one has
\eq
	\frac{1}{\tau}\int_0^\tau dt\,\sbkt{\bsphi(t),\Prnd\,\bsphi(t)}
	\le
	\frac{\tauB}{\tau}\,(1+2D^{-\kappa/5})\,(1+\varepsilon)
	\sim\frac{\tauB}{\tau},
\lb{main}
\en
for any normalized initial state $\bsphi(0)\in\calH$, and for any $\tau$ such that
\eq
0\le\frac{\tau}{\tauB}\le
\min\biggl\{D^{\kappa/5},\Bigl(\frac{D}{d}\Bigr)^{1/4}\biggr\}.
\lb{maintaurange}
\en
\end{theorem}
%%%%% THM %%%%%

\bigno
{\em Remarks:}\/
1.  With probability very close to one, the left-hand side of \rlb{main} converges to $d/D$ for extremely large $\tau$.
This fact was proved by von Neumann \cite{vonNeumann,GLMTZ10,GLTZ}.
%%%%%%%%%%%
\newremark
2.
Let us explain the motivation for the assumption \rlb{rEbD}.
In a macroscopic quantum system, the number of states (i.e., the number of $\alpha$ such that $E_\alpha\le E$) is well approximated by a function\footnote{%
The number of states generally behaves as $\Omega(E)\approx\exp[V\,\sigma(E/V)]$, where the entropy density $\sigma(\epsilon)$ is a strictly increasing convex function.
} $\Omega(E)$ which is smooth and rapidly increasing in $E$.
See, e.g., section~3.5 of \cite{Ruelle}.
Then the temperature $T$ at energy $U$ is written as
\eq
	\frac{1}{k_\mathrm{B}T}
	=\frac{\partial\log\Omega(E)}{\partial E}\Bigl|_{E=U}
	=\frac{\rho(U)}{\Omega(U)},
\en
where $\rho(E):=\Omega'(E)$ is the density of states.
By noting that $D=\Omega(U)-\Omega(U-\DU)\sim\Omega(U)$, where the approximate equality is generally quite accurate provided that $\DU\gg k_\mathrm{B}T$, and $\rho(E)$ is generally increasing in $E$, we find
\eq
	\rho(E)\le\rho(U)
	=\frac{1}{k_\mathrm{B} T} \,\Omega(U)
	\sim \frac{D}{k_\mathrm{B}T},
\en
for any $E\in[U-\DU,U]$,
which is our assumption \rlb{rEbD}.
%%%%%%%%%%%
\newremark
3.
The assumption $\DU\gg k_\mathrm{B}T$ in the previous remark is legitimate from a physical point of view.
In the unphysical situation with $\DU\ll k_\mathrm{B}T$, which is  easily realized mathematically, the assumption \rlb{rEbD} is no longer appropriate.
Here one expects the density of states $\rho(E)$ to be almost constant within the whole interval $[U-\DU,U]$.
Thus the assumption \rlb{rEbD} should be replaced by $\rho(E)\le D/\DU$.
Consequently the main inequality \rlb{main} of Theorem~\ref{t:main} is modified as
\eq
	\frac{1}{\tau}\int_0^\tau dt\,\sbkt{\bsphi(t),\Prnd\,\bsphi(t)}
	\lesssim\frac{h}{\DU}\frac{1}{\tau},
\lb{mainDU}
\en
which means that the decay is much slower than in the original setting.
See the end of section~\ref{s:physics} for an interpretation.
%%%%%%%%%%%
\newremark
4.
One may interpret our theorem, which is indeed valid for any $d$ with $1\le d\ll D$, as providing information about how fast states vary in a macroscopic quantum system.
We find that, with probability close to one, any state (including mixed states) in $\Hrnd$ escapes from $\Hrnd$ on the order of the Boltzmann time.

%%%%%%%%%%%%%%%%%%%%%%%%%%%%%%%%%%%%%%%%%%%
\subsection{Physical implications of the theorem}
\label{s:physics}
As we have briefly discussed in section~\ref{s:BG}, our main inequality \rlb{main} implies that any initial state of the system approaches thermal equilibrium within the time scale $\tau\gg\tauB$, provided that we regard $\Hrnd$ as a realization of the physical nonequilibrium subspace.

To see this rewrite \rlb{main} as 
\eq
\frac{1}{\tau}\int_0^\tau dt\,\sbkt{\bsphi(t),\Prnd\,\bsphi(t)}
\le
\varepsilon_1\varepsilon_2,
\lb{main2}
\en
where both $\varepsilon_1$ and  $\varepsilon_2$ are assumed to be small.
Then define the ``bad'' subset of $[0,\tau]$ by
\eq
\calB:=
\{t\in[0,\tau]\,\bigl|\,
\sbkt{\bsphi(t),\Prnd\,\bsphi(t)}\ge\varepsilon_1\}.
\en
From \rlb{main2} and the Markov inequality one readily finds that the ``total length'' (or the Lebesgue measure) $|\calB|$ of the subset $\calB$ satisfies
\eq
\frac{|\calB|}{\tau}\le\varepsilon_2,
\en
which means that the bad subset $\calB$ is a minority in the whole interval $[0,\tau]$.
In other words, we have $\sbkt{\bsphi(t),\Prnd\,\bsphi(t)}\le\varepsilon_1$ for a typical $t$ (randomly chosen) from $[0,\tau]$.
Since small $\sbkt{\bsphi(t),\Prnd\,\bsphi(t)}$ indicates that the state $\bsphi(t)$ is in equilibrium, we conclude that the system equilibrates within time $\tau$, no matter what the initial state $\bsphi(0)$ is.

Let us examine how large $\tau$ should be in a realistic setting.
Suppose that $T\approx300\rm\,K$, and hence $\tauB\approx1.6\times10^{-13}\rm\,s$.
Then even for $\tau\approx1\rm\,\mu s=10^{-6}\,s$, the right-hand side of \rlb{main} does not exceed $10^{-6}$.
One can safely take  $\varepsilon_1=\varepsilon_2=10^{-3}$, which means that the system is certainly in equilibrium after a micro-second.

As we have already discussed in Section~\ref{s:intro}, this mathematical conclusion is in a sharp contradiction with the empirical fact that there are so many nonequilibrium states which lasts for quite a long time\footnote{
We also note that the relaxation time should grow with the size of the system, while the Boltzmann time $\tauB$ is independent of the system size.
}.
The only reasonable resolution\footnote{%
The states in the energy shell $\calH$ are restricted in the sense that they are linear combinations of energy eigenstates corresponding to a  narrow range of energy.
We nevertheless expect that $\calH$ is large enough to contain many states which are sufficiently far from nonequilibrium.
} seems to be that realistic nonequilibrium subspaces form exceptions to the bound of the theorem, or, in other words, moderately slow decay observed in reality is not a typical property (if we assume random $\Hneq$).
See Section~\ref{s:intro} for a discussion about the implication of this finding to the method of appeal to typicality.

The atypicality may not be too surprising, especially after knowing about it.
Given the energy shell $\calH$, the nonequilibrium subspace $\Hneq$, in reality, is determined not in a random manner, but through the values of macroscopic quantities that we use to characterize the system.
Recall that many of the standard macroscopic quantities are expressed as a sum (or an integral) of locally conserved observables, and the Hamiltonian of a realistic system consists of more or less short-range interactions\footnote{%
In systems with short-range interactions, the relaxation time must grow with the system size because information propagates only with a finite speed.
In fact it is very likely that, in a suitable class of systems, one can use the Lieb-Robinson bound \cite{LiebRobinson} to prove lower bounds for the relaxation time which grows with the system size.
}.
This means that the corresponding subspace $\Hneq$ and the projection 
operator $\Pneq$ should be special.
It is likely, for example, that the commutator $[\hH,\Pneq]$ is smaller for realistic subspaces compared with randomly chosen ones. 

In order to fully understand the problem of the approach to equilibrium in macroscopic quantum systems, it may be essential to characterize realistic nonequilibrium subspaces, and to investigate the accompanying time scale.

Let us here give a crude picture based on the escape from a single state, which may be useful in understanding the nature of realistic nonequilibrium subspaces.
See \cite{Monnai} for a detailed study of the related problem.
Take an arbitrary state $\bsxi\in\calH$ and expand it as $\bsxi=\suma\xi_\alpha\psa$.
Assume that the coefficient $\xi_\alpha$ is negligible (in a certain rough sense) unless $\alpha$ is such that $E_\alpha\in[\tE,\tE+\DE]\subset[U-\DU,U]$ for some energy $\tE$ and energy width $\DE$.
We then take an initial state $\bsphi(0)\in\calH$ which is close to $\bsxi$, and examine how quickly the state escapes from the vicinity of $\bsxi$.
One readily finds that the overlap\footnote{%
Throughout the present paper we denote by $z^*$ the complex conjugate of  $z$.
}
\eq
	\sbkt{\bsphi(t),\hat{p}[\bsxi]\bsphi(t)}
	=\bigl|\sbkt{\bsxi,\bsphi(t)}\bigr|^2
	=\sumab(\xi_\alpha)^*\xi_\beta\,c_\alpha(c_\beta)^*\,
	e^{-i(\Eab)t/\hbar},
\en
where we expanded $\bsphi(0)$ as in \rlb{phi0exp},
changes (and hence decays) in the time scale of order $\taue:=h/\DE$.
This property is well-known as the ``uncertainty relation between time and energy''.

This observation suggests that a subspace $\Hneq$ may also be characterized by certain energy scale $\DE$, which is similarly related to the associated relaxation time.
This picture is true, at least for some examples, as we shall see now.

In the present setting of completely random $\Hneq$, we see that each $\bsxi^{(j)}$ (for $j=1,\ldots,d$) is characterized by the width (in the above sense) $\DE\approx k_\mathrm{B}T$.
This is  because, in a macroscopic system, most of the energy eigenvalues $E$ such that $U-\DU\le E\le U$ are found in the smaller range
with $U-\text{const}\,k_\mathrm{B}T\le E\le U$, where the constant is of $O(V^0)$. 
From this the conclusion that  the escape time coincides with the Boltzmann time does indeed  follow.

Two examples of $\Hneq$ in our previous work \cite{GHT13} are also consistent with the picture of escape time.
In Theorem~1, where one finds extremely slow decay, we have $\DE=\DU/d$, which means $\taue=h\,d/\DU$.
In Theorem~2, where one finds quick decay, we have $\DE\approx\DU$, which means $\taue\approx h/\DU$.

Remark~3 at the end of section~\ref{s:mainresult} also provides an example.
Here $\DU$ plays the role of the width $\DE$, which corresponds to the escape time $\taue=h/\DU$.
This is consistent with the bound \rlb{mainDU}.

This observation suggests that 
a realistic nonequilibrium
subspace $\Hneq$, determined through macroscopic quantities
that we use to characterize the system, is associated with a certain energy width $\DE$.
It is possible that the escape time $\taue$, which may take a reasonable value depending on the value of $\DE$, essentially determines the relaxation time.

Of course it is also likely that the above picture of the escape from a single state is too naive or has only limited applicability.
In some systems it may happen that $\bsphi(t)$ is ``trapped'' in the vicinity of $\Hneq$ in a more intricate manner.
In such a situation the relaxation time should also depend on the number of independent $\bsxi^{(j)}$'s that the state $\bsphi(t)$ should go through.

It is certainly an interesting challenge to examine these pictures in interacting many-body quantum systems and prove meaningful theorems.
For the moment we only have limited rigorous results in abstract and artificial settings.
In particular Theorem~1 of \cite{GHT13} treats examples with unphysically long relaxation time, and Theorem~\ref{t:main} of the present paper shows that a typical $\Hneq$ leads to unphysically short relaxation time.
The reality should lie in between these two extreme theorems, remaining to be understood.

%%%%%%%%%%%%%%%%%%%%%%%%%%%%%%%%%%%%%%%%%%%
%%%%%%%%%%%%%%%%%%%%%%%%%%%%%%%%%%%%%%%%%%%
\section{Preliminary considerations and the proof of Theorem~\protect\ref{t:main}}
\label{s:pre}

Here we make some preliminary considerations, and introduce important quantities including the two matrices $\sfS$ and $\sfQ$.
We shall prove Theorem~\ref{t:main} by using 
Propositions~\ref{t:main3cor} and \ref{c:mumax}.

Let $\bsphi(0)\in\calH$ be an arbitrary normalized initial state.
We expand $\bsphi(0)$ in the energy eigenstates as
\eq
	\bsphi(0)=\suma c_\alpha\psa,
	\lb{phi0exp}
\en
where the normalization implies $\suma|c_\alpha|^2=1$.

By recalling the time-evolution \rlb{phit}, we write the time average of the expectation value of $\Prnd$ as
\eqa
\frac{1}{\tau}\int_0^\tau dt\,\sbkt{\bsphi(t),\Prnd\,\bsphi(t)}
&=
\bbkt{\bsphi(0),
\Bigl\{\frac{1}{\tau}\int_0^\tau dt\,e^{i\hH t/\hbar}\Prnd\,e^{-i\hH t/\hbar}\Bigr\}\,
\bsphi(0)
}
\nl
&=\sumab(c_\alpha)^*Q_\ab\,c_\beta,
\lb{LTAandQ}
\ena
and we have introduced
\eq
Q_\ab:=\bbkt{\psa,
\Bigl\{\frac{1}{\tau}\int_0^\tau dt\,e^{i\hH t/\hbar}\Prnd\,e^{-i\hH t/\hbar}\Bigr\}\,
\psb
}.
\lb{Qdef}
\en
Let us write the ``matrix element'' of the projection $\Prnd$ as
\eq
P_\ab:=\sbkt{\psa,\Prnd\,\psb}=
\sum_{j=1}^d\xi^{(j)}_\alpha(\xi^{(j)}_\beta)^*,
\lb{Pab}
\en
where we used \rlb{Pdef}, and expanded the (random) states as $\bsxi^{(j)}=\suma\xi^{(j)}_\alpha\psa$.
Since $\bsxi^{(j)}$ is normalized we have $\suma|\xi^{(j)}_\alpha|^2=1$.
Then by using \rlb{Hp=Ep}, we can rewrite \rlb{Qdef} as
\eq
Q_\ab=\frac{1}{\tau}\int_0^\tau dt\,e^{iE_\alpha t/\hbar}\,
P_\ab\,e^{-iE_\beta t/\hbar}
=P_\ab\,S_\ab,
\lb{QPS}
\en
with
\eq
S_\ab:=
\frac{1}{\tau}\int_0^\tau dt\,e^{i(E_\alpha-E_\beta)t/\hbar}
=
\begin{cases}
1,&\text{if $E_\alpha=E_\beta$};\\
\dfrac{e^{i\tau(E_\alpha-E_\beta)/\hbar}-1}{i\tau(E_\alpha-E_\beta)/\hbar},&
\text{if $E_\alpha\ne E_\beta$.}
\end{cases}
\lb{Sdef}
\en
We then define $D\times D$ matrices by $\sfP=(P_\ab)_{\alpha,\beta\in\oD}$, $\sfS=(S_\ab)_{\alpha,\beta\in\oD}$, and $\sfQ=(Q_\ab)_{\alpha,\beta\in\oD}$.
The matrix $\sfQ$ is the Hadamard product of $\sfP$ and $\sfS$.

Note that $\sfS$ is positive semi-definite 
as, for any $(c_\alpha)_{\alpha\in\oD}$, we have
\eq
	\sumab (c_\alpha)^*\,S_\ab\,c_\beta
	=\frac{1}{\tau}\int_0^\tau dt\,
	\biggl|\suma c_\alpha\,e^{-iE_\alpha t/\hbar}\biggl|^2\ge0.
\en
We also note that $\sfQ$ is hermitian since both $\sfP$ and $\sfS$ are hermitian\footnote{%
It follows from \rlb{Qdef} that $\sfQ$ is positive semi-definite.
See also the proof of Proposition~\ref{t:main3cor}, especially \rlb{Qpos}.
}.

By $\mumax$ and $\lamax$ we denote the maximum eigenvalues of $\sfS$ and $\sfQ$, respectively.
Note that \rlb{LTAandQ} implies
\eq
\frac{1}{\tau}\int_0^\tau dt\,\sbkt{\bsphi(t),\Prnd\,\bsphi(t)}
=\sumab(c_\alpha)^*Q_\ab\,c_\beta
\le\lamax
\lb{LTAandlamax}
\en
where we noted that $\suma|c_\alpha|^2=1$.
Since the left-hand side of \rlb{LTAandQ} never exceeds 1, we find that $\lamax\le1$.

\bigno
{\em Remark:}\/
It is useful to note that, for each (fixed) random unitary transformation $\hat{U}$, the maximum eigenvalue $\lamax$ is nondecreasing in $d$.
To see this, we write the $d$-dependences of the matrices explicitly, and note that \rlb{Qdef} implies
$Q_\ab^{(d')}-Q_\ab^{(d)}=\bbkt{\psa,
\bigl\{\tau^{-1}\int_0^\tau dt\,e^{i\hH t/\hbar}(\Prnd^{(d')}-\Prnd^{(d)})\,e^{-i\hH t/\hbar}\bigr\}\,\psb}$.
Suppose $d'\ge d$.
Since $\Prnd^{(d')}-\Prnd^{(d)}$ is a projection, we see that $\sfQ^{(d')}-\sfQ^{(d)}$ is positive semi-definite.

\bigskip

Mathematically speaking our most important result is the following proposition which roughly says that one typically has $\lamax\lesssim\mumax/D$ (provided that $\mumax$ is large enough).
We shall here state the proposition for general positive semi-definite matrices since this mathematical result itself may be of some interest from the point of view of random matrices.
The proposition is proved in section~\ref{s:proofofmain2} by using Proposition~\ref{t:main3}.

%%%%% PROP %%%%%
\begin{proposition}
\label{t:main3cor}
Suppose that the dimensions $D$ and $d$, and an integer 
$n\ge2$ satisfy $n^4d\le 2D$. 
Let $\sfA=(A_\ab)_{\alpha,\beta\in\oD}$ be a $D\times D$ positive 
semi-definite matrix 
whose maximum eigenvalue is $\mumax$.
Take a constant $\bmu$ such that
\eq
	\mu_{\max} \leq \bmu, \qquad 
	\text{ and } \qquad 
	\Bigl ( \frac{d}{D} \Bigr )^{1/4} \leq\frac{\bmu}{D\,\bar{A}} 
	,
\lb{Acond-cor1}
\en
where $\bar{A}:= \max_{1\leq \alpha\leq D} A_{\alpha \alpha}$ is assumed to be nonvanishing.
Define another $D\times D$ matrix  $\sfB=(B_\ab)_{\alpha,\beta\in\oD}$ by the Hadamard product
\eq
	B_\ab=A_\ab P_\ab\qquad (\alpha,\beta\in\oD),
\lb{BAP2}
\en
where $P_\ab$ are the matrix elements of the projection operator onto a random $d$-dimensional subspace as in \rlb{Pdef} and \rlb{Pab}. 

Then for any $\varepsilon>0$, the maximum eigenvalue $\lamax$ of the matrix $\sfB$ satisfies\footnote{%
When we apply the proposition, we let $n$ be large to make the right-hand side of \rlb{PlamaxGeneral1} small.
}
\eq
	\prob\Bigl[
	\lamax \geq (1+\varepsilon)\, \frac{\bmu}{D}
	\Bigr]
	\leq 
	\frac{2 e^3 d D^{1/4}} {(1+\varepsilon)^{n}} \, 
\lb{PlamaxGeneral1}
\en
where the probability is with respect to the random choice of the $d$-dimensional subspace.
\end{proposition}
%%%%% PROP %%%%%

\noindent
{\em Remark:}\/
There is a similar result which holds for hermitian (and not necessarily positive-semidefinite) matrix $\sfA$.
Decompose $\sfA$ into its positive and negative parts\footnote{%
	Let $(u^{(\ell)}_\alpha)_{\alpha\in\oD}$ and $\mu_\ell$ 
	($\ell\in\oD$) be the eigenvectors and the corresponding 
	eigenvalues of $\sfA$, respectively.
	We then define 
	$A^{(+)}_{\alpha \beta} := \sum_{\ell: \mu_\ell \geq 0} 
	\mu_\ell \,u^{(\ell)}_\alpha ( u^{(\ell)}_\beta)^*$, and 
	$A^{(-)}_{\alpha \beta} := \sum_{\ell:\mu_\ell < 0} 
	\mu_\ell \,u^{(\ell)}_\alpha ( u^{(\ell)}_\beta)^*$, 
	where $\sfA^{(+)}$ is positive semi-definite, and  
	$\sfA^{(-)}$ is negative definite.} 
as $\sfA=\sfA^{(+)}+\sfA^{(-)}$.
Then $\sfB$ is also decomposed as $\sfB = \sfB^{(+)}+ \sfB^{(-)}$, where $\sfB^{(\pm)}$ denotes the Hadamard product of $\sfA^{(\pm)}$ and $\sfP$.
Since
\eqalign 
	\lamax 
	& \leq \bigl ( \text{max eigenvalue of } \sfB^{(+)} 
	\bigr ) 
	+ \bigl ( \text{max eigenvalue of } \sfB^{(-)} 
	\bigr ) 
	\nnb
	& 
	\leq \bigl ( \text{max eigenvalue of } \sfB^{(+)} 
	\bigr ),
\enalign
we only need to bound the maximum eigenvalue of $\sfB^{(+)}$.
This can be done by using Proposition~\ref{t:main3cor} for $\sfA^{(+)}$ and $\sfB^{(+)}$.

\bigskip

We shall apply Proposition~\ref{t:main3cor} by setting $\sfA=\sfS$.
This means that we need an upper bound $\bmu$ for the maximum eigenvalue $\mumax$ of $\sfS$.
We make use of the following upper bound, which indeed is almost optimal\footnote{
By using the variational argument described in \cite{GHT13B}, one can prove a lower bound which has the same asymptotic behavior (as $D\up\infty$) as the upper bound.
}.
%%%%% PROP %%%%%
\begin{proposition}
\label{c:mumax}
Assume \rlb{eta-def}
and \rlb{rEbD} for the density of states $\rho(E)$.
Then the maximum eigenvalue $\mumax$ of the matrix $\sfS$ satisfies
\eq
	\mumax\le\bmu:=\frac{\tauB}{\tau}D 
	\, \frac{1 + D^{-\kappa/5}/2}{1-D^{-\kappa/5}}
	\sim \frac{\tauB}{\tau}D,
\lb{mumax}
\en
for any $\tau$ such that $0<\tau\le D^{\kappa/5}\,\tauB$.
\end{proposition}
%%%%% PROP %%%%%
The proposition will be proved at the end of the present section.

\bigno
{\em Remark:}\/
The proposition may have applications in various problems of quantum mechanical time evolution.
For example the function $\phi(t):=D^{-1}\suma e^{itE_\alpha/\hbar}$ plays a fundamental role in the work by Cramer \cite{Cramer}.
We can bound the time average of  $|\phi(t)|^2$ as
\eq
	\frac{1}{\tau}\int_0^\tau dt\,|\phi(t)|^2
	=D^{-1}\sumab D^{-1/2}S_\ab D^{-1/2}
	\le\frac{\mumax}{D}\le\frac{\bmu}{D},
\en
which may be useful in extending the results\footnote{%
It is likely that, for the class of models (with $N$ sites) in section~3.2 of \cite{Cramer}, the bound \rlb{rEbD} for the density of states is valid with $\tilde{\beta}=\text{const}\times N$.
} in \cite{Cramer}.

\bigskip

We are now ready to verify that our main theorem follows from these propositions.

\bigno
{\em Proof  of Theorem~\ref{t:main}, assuming Propositions~\ref{t:main3cor} and \ref{c:mumax}:}\/
We shall set $\sfA=\sfS$. 
We have already remarked that $\sfS$ is positive semi-definite. 
Also $\bar{A}=1$ 
because of the definition \rlb{Sdef}.

We note that the condition \rlb{maintaurange} in the theorem implies $\tau/\tauB\le(D/d)^{1/4}$, which shows that $\bmu$ given by \rlb{mumax} satisfies $\bmu/D\ge(d/D)^{1/4}$.
This justifies \rlb{Acond-cor1}.

For given $\varepsilon>0$, we let $n$ be the smallest integer such that $2e^3dD^{1/4}/(1+\varepsilon)^n\le d/D$.
Suppose that $n^4d\le2D$.
Then \rlb{PlamaxGeneral1} along with \rlb{LTAandlamax} implies the desired \rlb{main} because $\sfB$ corresponds to $\sfQ$.

In \rlb{main}, we have replaced $\bmu$ of 
\rlb{mumax} by its upper bound 
$({\tauB}/{\tau})D(1+2D^{-\kappa/5})$, which is justified 
because $D$ is sufficiently large as is explained below.

Let us  examine the condition that $n^4d\le 2D$.
Since 
$n\le\{\log(2e^3D^{5/4})/\log(1+\varepsilon)\}+1$, the condition $n^4d\le 2D$ is guaranteed if 
\rlb{main-cond-on-d_and_D} is satisfied. 
To ensure that there is a positive $d$ which satisfies the inequality 
\rlb{main-cond-on-d_and_D}, $D$ must be sufficiently large to satisfy 
$[\{\log(2e^3D^{5/4})/\log(1+\varepsilon)\}+1]^4\le D$.
If one chooses $\varepsilon=0.01$, it suffices to set, e.g., $D=10^{50}$ and $d=10^{30}$. 
For a typical value of $D$ such as $D \approx \exp(10^{20})$,  the condition \rlb{main-cond-on-d_and_D} is definitely satisfied even for $d$ as large as, say, $d \approx \exp(0.9999\times10^{20})$.
\qedm

\bigskip

Let us end the section by proving the upper bound on the maximum eigenvalue.

\bigno
{\em Proof of Proposition~\ref{c:mumax}:}\/
Let $u>0$.
We define $S^{(u)}_\ab$ by replacing $\tau$ in \rlb{Sdef} by $u$.
We also define
\eq
R^{(u)}_\ab:=u\,e^{-iuE_\alpha/(2\hbar)}\,S^{(u)}_\ab\,e^{iuE_\beta/(2\hbar)}.
\en
Note that the matrices $(S^{(u)}_\ab)_{\alpha,\beta\in\oD}$ and $(e^{-iuE_\alpha/(2\hbar)}\,S^{(u)}_\ab\,e^{iuE_\beta/(2\hbar)})_{\alpha,\beta\in\oD}$ have exactly the same eigenvalues since they are related with each other by a trivial unitary transformation\footnote{
The latter is obtained by replacing the range of integration in \rlb{Sdef} by $[-\tau/2,\tau/2]$.
}.
Thus the maximum eigenvalue of the matrix $(R^{(\tau)}_\ab)_{\alpha,\beta\in\oD}$ is equal to $\tau\,\mumax$.

From \rlb{Sdef} we find
\eq
R^{(u)}_\ab
=\int_{-u/2}^{u/2}dt\,e^{i(\Eab)t/\hbar}
=
\begin{cases}
u,&\text{if $E_\alpha=E_\beta$};\\
\dfrac{\sin\bigl[(\Eab)u/(2\hbar)\bigr]}{(\Eab)/(2\hbar)},
&
\text{if $E_\alpha\ne E_\beta$.}
\end{cases}\lb{Rdef}
\en
which means that $(R^{(\tau)}_\ab)_{\alpha,\beta\in\oD}$ is a real symmetric matrix.
Thus the maximum eigenvalue can be written as
\eq
\tau\,\mumax=\suptwo{x_1,\ldots,x_D\in\bbR}{\bigl(\suma(x_\alpha)^2=1\bigr)}
\sumab x_\alpha R^{(\tau)}_\ab\,x_\beta.
\lb{mumaxvar}
\en
This variational problem is not yet easy to treat since $R^{(\tau)}_\ab$ has an oscillating sign.
We overcome this difficulty by performing an extra integration in time.

Fix arbitrary $x_1,\ldots,x_D\in\bbR$ such that $\suma(x_\alpha)^2=1$.
From the integral representation in \rlb{Rdef}, we get
\eq
\sumab x_\alpha R^{(u)}_\ab\,x_\beta
=\int_{-u/2}^{u/2}dt\,\Bigl|\suma e^{-iE_\alpha t/\hbar}\,x_\alpha\Bigr|^2,
\en
which implies that the quantity on 
the left-hand side is nonnegative and increasing in $u>0$.
Thus, for any $0<\tau<\taum$, we have
\eqa
\sumab x_\alpha R^{(\tau)}_\ab\,x_\beta
&\le\frac{1}{\taum-\tau}\int_\tau^{\taum} du\,\sumab x_\alpha R^{(u)}_\ab\,x_\beta
\nl&
\le\frac{1}{\taum-\tau}\int_0^{\taum} du\,\sumab x_\alpha R^{(u)}_\ab\,x_\beta
=\sumab x_\alpha\,I_\ab\,x_\beta,
\lb{xRx}
\ena
where
\eq
I_\ab:=\frac{1}{\taum-\tau}\int_0^{\taum} du\,R^{(u)}_\ab
=\frac{(\taum)^2}{2(\taum-\tau)}\,\,g\Bigl(\frac{(\Eab)\,\taum}{4\hbar}\Bigr),
\en
with $g(x):=(\sin x/x)^2$.
Remarkably $I_\ab$ is nonnegative.

By noting that $2x_\alpha x_\beta\le(x_\alpha)^2+(x_\beta)^2$ and $I_\ab=I_{\beta\alpha}\ge0$, we find
\eqa
\sumab x_\alpha\,I_\ab\,x_\beta&\le
\frac{1}{2}\sumab\bigl\{(x_\alpha)^2 I_\ab+I_\ab(x_\beta)^2\bigr\}
=\sumab I_\ab(x_\beta)^2
\nl&
\le\Bigl(\sup_{\beta\in\oD}\suma I_\ab\Bigr)\sum_{\beta=1}^D(x_\beta)^2
=\sup_{\beta}\suma I_\ab.
\ena
Recalling \rlb{mumaxvar} and \rlb{xRx}, we get our main bound
\eq
\tau\,\mumax\le \sup_{\beta}\suma I_\ab.
\lb{mumaxI}
\en
By replacing the sum on the right-hand side of \rlb{mumaxI} by an integral according to \rlb{sumtoint}, we have
\eqa
\tau\,\mumax&\le
\frac{(\taum)^2}{2(\taum-\tau)}
\,\sup_\beta\biggl\{
\int_{U-\DU-\eta}^{U+\eta}dE\,\rho(E)\,g\Bigl(\frac{(E-E_\beta)\,\taum}{4\hbar}\Bigr)
+D^{1-\kappa}\frac{\taum\,\bar{g}}{4\hbar\tbeta}
\biggr\},
\itext{%
where $\bar{g}:=\sup_x|g'(x)|\sim0.5402$.
By using \rlb{rEbD}, the integral is bounded as
}
&\le
\frac{(\taum)^2}{2(\taum-\tau)}
\,\sup_\beta\biggl\{
D\tbeta\int_{U-\DU-\eta}^{U+\eta}dE\,g\Bigl(\frac{(E-E_\beta)\,\taum}{4\hbar}\Bigr)
+D^{1-\kappa}\frac{\pi\taum\,\bar{g}}{2\tauB}
\biggr\},
\nl&
\le
\frac{(\taum)^2}{2(\taum-\tau)}
\,\biggl\{
D\tbeta\int_{-\infty}^\infty dE\,g\Bigl(\frac{E\,\taum}{4\hbar}\Bigr)
+D^{1-\kappa}\frac{\pi\taum\,\bar{g}}{2\tauB}
\biggr\},
\itext{%
where we also used the definition \rlb{tauB} of $\tauB$ to rewrite the error term.
Recalling that $\int_{-\infty}^\infty dx\,g(x)=\pi$, the integration can be evaluated to give
}
&\le \frac{\taum}{\taum-\tau}D\,\tauB
\Bigl(
1+\frac{\pi\bar{g}}{4}\Bigl(\frac{\taum}{\tauB}\Bigr)^2D^{-\kappa}
\Bigr).
\ena
We finally choose $\taum=D^{2\kappa/5}\tauB$, and use the upper bound $\tau\le D^{\kappa/5}\,\tauB$ to get
\eq
	\mumax\le\frac{1}{\tau}\frac{1}{1-D^{-\kappa/5}}D\,\tauB
	\Bigl(
	1+\frac{\pi\bar{g}}{4}D^{-\kappa/5}
	\Bigr)
	\le \frac{\tauB}{\tau}D \, 
	\frac{1+D^{-\kappa/5}/2}{1-D^{-\kappa/5}},
\en
which is the desired upper bound.\qedm

%%%%%%%%%%%%%%%%%%%%%%%%%%%%%%%%%%%%%%%%%%%
%%%%%%%%%%%%%%%%%%%%%%%%%%%%%%%%%%%%%%%%%%%
\section{Main proposition and its proof}
\label{s:SQ}
In the present section, which is the mathematical core of the present work, we state and prove Proposition~\ref{t:main3} about general matrices.
Proposition~\ref{t:main3cor} is proved rather easily from Proposition~\ref{t:main3}.

%%%%%%%%%%%%%%%%%%%%%%%%%%%%%%%%%%%%%%%%%%%
\subsection{Proof of Proposition~\protect\ref{t:main3cor}}
\label{s:proofofmain2}

Let us state our main proposition.
It deals with a matrix $\sfA$ (with complex elements) which is not necessarily positive semi-definite, but satisfies a special condition \rlb{Acond}. 
%%%%% PROP %%%%%
\begin{proposition}
\label{t:main3}
Suppose that the dimensions $D$ and $d$, an integer $n\ge2$, and 
some $\delta>0$ satisfy 
$n^4d\le 2D$ and $n^{2+(4/\delta)}d\le D$.
Let $\sfA=(A_\ab)_{\alpha,\beta\in\oD}$ be a $D\times D$ 
matrix which satisfies
\eq
|	(\sfA^m)_\ab|\le\bmu^{m-1}, \qquad (\alpha, \beta \in \{1, 2, 
\ldots, n\})
\lb{Acond}
\en
for any positive integer $m$ with an $m$-independent 
constant $\bmu>0$.

Define another $D\times D$ matrix  $\sfB=(B_\ab)_{\alpha,\beta\in\oD}$ by the Hadamard product
\eq
B_\ab=A_\ab P_\ab,
\lb{BAP}
\en
where $P_\ab$ are the matrix elements of the projection operator onto a random $d$-dimensional subspace as in \rlb{Pdef} and \rlb{Pab}.
Then we have\footnote{%
We write $a\vee b=\max\{a,b\}$.
}
\eq
	\Bigl|\Av\bigl[\tr[\sfB^n]\bigr]\Bigr|
	\le 2e^2d
	\biggl [ 
	\Bigl ( \frac{\bmu}{D} \Bigr )^{n-1} 
	+ (e-1) \, \biggl \{ \frac{\bmu}{D} \vee 
	\Bigl ( \frac{d}{D} \Bigr )^{1/(2+\delta)}\biggr \}^{n}\, 
	\biggr ],
	\lb{trBn}
\en
where the expectation is with respect to the random choice of the $d$-dimensional subspace.
\end{proposition}
%%%%% PROP %%%%%

\bigno
{\em Proof of Proposition~\ref{t:main3cor}, assuming Proposition~\ref{t:main3}:}\/
We will prove the proposition assuming $\bar{A} = 1$. 
The case with $\bar{A} \neq 1$ can be 
reduced to this case by considering matrices 
$\tilde{\sfA}$ and $\tilde{\sfB}$, whose elements are defined as
$\tilde{A}_{\alpha \beta} = A_{\alpha \beta} / \bar{A}$ and 
$\tilde{B}_{\alpha \beta} = B_{\alpha \beta} / \bar{A}$, respectively.

We begin by checking various conditions in Proposition~\ref{t:main3}. 
The condition $n^4d\le 2D$ in Proposition~\ref{t:main3} is also 
assumed in Proposition~\ref{t:main3cor}.
We set $\delta=2$.
Then the condition $n^{2+(4/\delta)}d\le D$ is satisfied since we have $n^4d\le2D$.

We then verify the condition \rlb{Acond}.
Recall that now $\sfA$ is assumed to be positive semi-definite.
Let $(u^{(\ell)}_\alpha)_{\alpha\in\oD}$ and $\mu_\ell\ge0$ 
($\ell\in\oD$) be the eigenvectors and the corresponding 
eigenvalues of $\sfA$, respectively. 
We can write 
$A_\ab=\sum_{\ell=1}^{D} \mu_\ell \,u^{(\ell)}_\alpha( u^{(\ell)}_\beta)^*$.
Then we note for $m\ge1$ that
\eq
\bigl|(\sfA^m)_\ab\bigr|=\Bigl|\sum_{\ell=1}^{D} (\mu_\ell)^m u^{(\ell)}_\alpha( u^{(\ell)}_\beta)^*\Bigr|
\le\sum_{\ell=1}^{D} (\mu_\ell)^m\frac{1}{2}
\bigl\{|u^{(\ell)}_\alpha|^2+|u^{(\ell)}_\beta|^2\bigr\}.
\lb{Am1}
\en
We then observe that
\eq
\sum_{\ell=1}^{D} (\mu_\ell)^m|u^{(\ell)}_\alpha|^2
\le\bmu^{m-1}\sum_{\ell=1}^{D} \mu_\ell\,
u^{(\ell)}_\alpha( u^{(\ell)}_\alpha)^*
=\bmu^{m-1}\,A_{\alpha\alpha}\le\bmu^{m-1},
\lb{Am2}
\en
where we used \rlb{Acond-cor1} and the fact that $\bar{A} =1$. 
By substituting \rlb{Am2} into \rlb{Am1}, we get the desired bound \rlb{Acond}.

We next note that $\sfB$ is positive semi-definite. 
This is guaranteed by Schur's theorem, which says that the Hadamard 
product of two positive semi-definite matrices is 
also positive semi-definite, but let us give an elementary proof. 
Take an arbitrary $(x_\alpha)_{\alpha\in\oD}$ and observe that
\eqa
\sumab(x_\alpha)^*B_\ab\,x_\beta
&=\sumab(x_\alpha)^*P_\ab\,A_\ab\,x_\beta
=\sumab\sum_{j=1}^D(x_\alpha)^*\xi^{(j)}_\alpha(\xi^{(j)}_\beta)^*A_\ab\,x_\beta
\nl
&=\sum_{j=1}^D\sumab(y^{(j)}_\alpha)^*A_\ab\,y^{(j)}_\beta\ge0,
\lb{Qpos}
\ena
where we used \rlb{BAP2}, \rlb{Pab}, and set $y^{(j)}_\alpha=x_\alpha(\xi^{(j)}_\alpha)^*$.
We also used the fact that $\sfA$ is positive semi-definite.

Now positive semi-definiteness of $\sfB$ implies that
\eq
\tr[\sfB^n]=\sum_{\ell=1}^D(\lambda_\ell)^n\ge(\lamax)^n,
\en
where $\lambda_\ell\ge0$ with $\ell\in\oD$ are 
the eigenvalues of $\sfB$.

Since we have \rlb{Acond-cor1}, 
the main inequality \rlb{trBn} implies
\eqa 
	\Av[(\lamax)^n]\le \Av\bigl[\tr[\sfB^n]\bigr]
	& \le
	2e^2d
	\biggl \{
	\Bigl ( \frac{\bmu}{D} \Bigr )^{n-1} \!\!\!\!
	+ (e-1) \, \biggl(\frac{\bmu}{D}\biggr)^{n} 
	\biggr \}
	\nl 
	& \leq 2e^3d \Bigl(\frac{\bmu}{D}\Bigr)^{n-1} 
	\Bigl ( 1 \, \vee \, \frac{\bmu}{D} \Bigr  ) .
%	\nl
\ena
By using the Markov inequality, we find for any $\varepsilon>0$ that
\eqa
	& \prob\Bigl[
	\lamax \geq (1+\varepsilon)\, \frac{\bmu}{D} 
	\Bigr]
	= 
	\prob\Bigl[
	\bigl(\lamax\bigr)^{n} \geq  
	\Bigl\{ (1+\varepsilon)\, \frac{\bmu}{D} \Bigr\}^{n}
	\Bigr]
	\leq \frac{\Av \bigl [ \bigl( \lamax \bigr)^{n} \bigr ]}
	{\Bigl\{ (1+\varepsilon)\, \dfrac{\bmu}{D} \Bigr\}^{n}}
	\nl 
	& \qquad 
	\leq \frac{2 e^3 d  
	\Bigl( \dfrac{\bmu}{D} \Bigr)^{n-1} }
	{\Bigl\{ (1+\varepsilon)\, \dfrac{\bmu}{D} \Bigr\}^{n}} \, 
	\Bigl ( 1 \, \vee \, \frac{\bmu}{D} \Bigr  )
	= \frac{2 e^3 d} {(1+\varepsilon)^{n}} \, 
	\Bigl ( \frac{D}{\bmu} \, \vee \, 1 \Bigr  )
	\le
	\frac{2 e^3 d} {(1+\varepsilon)^{n}} \, D^{1/4}
	, 
\ena
where, for $\bmu \leq D$, we used \rlb{Acond-cor1} and $d\ge1$ to 
get the final bound.
\qedm

%%%%%%%%%%%%%%%%%%%%%%%%%%%%%%%%%
\subsection{Proof of Proposition~\protect\ref{t:main3}}
Now we shall prove Proposition~\ref{t:main3}.
The exact expression \rlb{UN-int.1} of Collins for integrals over the Haar measure of the unitary group plays a fundamental role in the proof.
We make use of some properties of the symmetric group, which are summarized in the Appendix~\ref{app-gp}.

%%%%%%%%%%%%%%%%%%%%%%%%%%
\subsubsection{Integration of $d$-dimensional random subspaces}
We shall examine the expectation value $\Av[\tr[\sfB^n]]$, and rewrite it into a form suitable for further evaluation.
From \rlb{BAP}, we have
\eqa
	\Av \bigl [ \tr [\sfB^n] \bigr ]
	& = \Av \Biggl[\,\,\sum_{\alpha_1, \ldots, \alpha_n=1}^D
	B_{\alpha_1 \alpha_2} \, B_{\alpha_2 \alpha_3} \, \cdots \,  
	B_{\alpha_n \alpha_1} 
	\Biggr ]
	\nl
	& = \sum_{\alpha_1, \ldots, \alpha_n=1}^D 
	A_{\alpha_1\alpha_2} \, A_{\alpha_2 \alpha_3} \, \cdots \,  
	A_{\alpha_n \alpha_1} \, 
	\Av \bigl [ 
	P_{\alpha_1 \alpha_2} \, P_{\alpha_2 \alpha_3} \, \cdots \,  
	P_{\alpha_n \alpha_1} 
	\bigr ] 
	.
	\lb{TrBn1}
\ena
By writing the projection matrix in terms of the unit vectors as in \rlb{Pab}, we can write the expectation value in the sum as
\eqa
	\Av \bigl [  
	P_{\alpha_1 \alpha_2} \, &P_{\alpha_2 \alpha_3} \, \cdots \,  
	P_{\alpha_n \alpha_1} 
	\bigr ]
	\nl&
	= \sum_{j_1, \ldots, j_n=1}^d 
	\Av \bigl [\,
	\xi^{(j_1)}_{\alpha_1} \,(\xi^{(j_1)}_{\alpha_2})^* \, 
	\xi^{(j_2)}_{\alpha_2} \,(\xi^{(j_2)}_{\alpha_3})^* \, 
	\cdots 
	\xi^{(j_{n-1})}_{\alpha_{n-1}} \,
	(\xi^{(j_{n-1})}_{\alpha_{n}})^* \, 
	\xi^{(j_n)}_{\alpha_{n}} \,
	(\xi^{(j_n)}_{\alpha_{1}})^* \, 
	\bigr ] 
	.
	\lb{aveP.11}
\ena

Now we need to evaluate 
$\Av \bigl [\,
	\xi^{(j_1)}_{\alpha_1} \,(\xi^{(j_1)}_{\alpha_2})^* \, 
	\cdots 
	\xi^{(j_n)}_{\alpha_{n}} \,
	(\xi^{(j_n)}_{\alpha_{1}})^* \, 
	\bigr ] $.
Let us first be heuristic and present a rough estimate.
Although we still know very little about this expectation value, it is apparent from the normalization and the symmetry that
\eq
\Av\bigl[\,\xi^{(j)}_{\alpha} \, (\xi^{(j)}_{\alpha'})^* \bigr]
=\frac{1}{D}\,\delta_{\alpha,\alpha'}
\lb{free1}
\en
holds.
It is expected that, in a crude approximation, one may treat 
\newline$(\xi^{(1)}_{\alpha})_{\alpha\in\oD},\ldots,(\xi^{(d)}_{\alpha})_{\alpha\in\oD}$ as independent random vectors each of which distributed uniformly on the unit sphere in $\mathbb{C}^D$.
Then \rlb{free1} is automatically satisfied.
Of course the different vectors are not necessarily orthogonal with each other in this approximation, but they are almost orthogonal with probability close to one provided that $d\ll D$.
In fact it was shown by Weingarten in his pioneering work on the group integrals that this crude approximation gives the leading orders of the large-$D$ limits of certain expectation values \cite{Weingarten}.

In this ``first-order approximation'' (and assuming that for most of the terms in \rlb{aveP.11} the $j_i$ ($i=1,\ldots,n$) are all different) the relevant expectation becomes
\eq
\Av \bigl [\,
	\xi^{(j_1)}_{\alpha_1} \,(\xi^{(j_1)}_{\alpha_2})^* \, 
	\xi^{(j_2)}_{\alpha_2} \,(\xi^{(j_2)}_{\alpha_3})^* \, 
	\cdots 
	\xi^{(j_n)}_{\alpha_{n}} \,
	(\xi^{(j_n)}_{\alpha_{1}})^* \, 
	\bigr ] 
	\sim
	\frac{1}{D^n}\,\prod_{s=1}^n\delta_{\alpha_s,\alpha_{s+1}}.
	\lb{free2}
\en
Here (and in what follows) we identify $\alpha_{n+1}$ with $\alpha_1$.
Substituted into \rlb{aveP.11}, this approximation yields
\eq
	\Av \bigl [  
	P_{\alpha_1 \alpha_2} \, P_{\alpha_2 \alpha_3} \, \cdots \,  
	P_{\alpha_n \alpha_1} 
	\bigr ]
	\sim \Bigl(\frac{d}{D}\Bigr)^n\,
	\prod_{s=1}^n\delta_{\alpha_s,\alpha_{s+1}}.
	\lb{free3}
\en
Note that the factor $\prod_{s=1}^n\delta_{\alpha_s,\alpha_{s+1}}$ imposes the constraint that $\alpha_1,\ldots,\alpha_n$ must be all identical.
Thus, recalling \rlb{TrBn1}, the present approximation gives
\eq
\Av \bigl [ \tr [\sfB^n] \bigr ]\stackrel{\text{?}}{\sim}
\Bigl(\frac{d}{D}\Bigr)^n\sum_{\alpha=1}^D(A_{\alpha\alpha})^n.
\lb{free4}
\en
This estimate, with the assumption \rlb{Acond}, implies
\eq
\Bigl|\Av \bigl [ \tr [\sfB^n] \bigr ]\Bigr|\stackrel{\text{?}}{\lesssim}
\Bigl(\frac{d}{D}\Bigr)^n\sum_{\alpha=1}^D|A_{\alpha\alpha}|^n
\le D\,\Bigl(\frac{d}{D}\Bigr)^n,
\lb{free5}
\en
which, however, turns out {\em not}\/ to be the major contribution to $\Av[\tr[\sfB^n]]$.
This is most easily seen by noticing that we have $A_{\alpha\alpha}=S_{\alpha\alpha}=1$ in our original problem.
Then the right-hand side of \rlb{free4} is simply $D\,(d/D)^n$, which is independent of $\tau$.
Although it is true that the approximations \rlb{free2} or \rlb{free3} gives the main term of the expectation value, the coupling with $A_\ab$ in \rlb{TrBn1} suppresses its contribution.

To find another contribution to $\Av[\tr[\sfB^n]]$, we set $j_1=\cdots=j_n=j$ in the expectation value
$\Av \bigl [\,
	\xi^{(j_1)}_{\alpha_1} \,(\xi^{(j_1)}_{\alpha_2})^* \, 
	\cdots 
	\xi^{(j_n)}_{\alpha_{n}} \,
	(\xi^{(j_n)}_{\alpha_{1}})^* \, 
	\bigr ] $.
We then find
\eq
	\Av \bigl [\,
	\xi^{(j)}_{\alpha_1} \,(\xi^{(j)}_{\alpha_2})^* \, 
	\xi^{(j)}_{\alpha_2} \,(\xi^{(j)}_{\alpha_3})^* \, 
	\cdots 
	\xi^{(j)}_{\alpha_{n}} \,
	(\xi^{(j)}_{\alpha_{1}})^* \, 
	\bigr ] 
	=
	\Av \bigl [\,
	|\xi^{(j)}_{\alpha_1}|^2\,|\xi^{(j)}_{\alpha_2}|^2
	\cdots|\xi^{(j)}_{\alpha_n}|^2
	\bigr]
	\sim\frac{1}{D^n},
	\lb{free6}
\en
where we assumed for simplicity that all $\alpha_1,\ldots,\alpha_n$ are distinct, and used the fact that for any $j$ the coefficients $\xi_\alpha^{(j)}$ (with $\alpha=1,\ldots,D$) of the random vector $\bsxi^{(j)}$ can be treated as almost independent random variables.
Assuming that \rlb{free2} and \rlb{free6} give the dominant contributions, we find from \rlb{aveP.11} that
\eq
	\Av \bigl [  
	P_{\alpha_1 \alpha_2} \, P_{\alpha_2 \alpha_3} \, \cdots \,  
	P_{\alpha_n \alpha_1} 
	\bigr ]
	\sim \Bigl(\frac{d}{D}\Bigr)^n\,
	\prod_{s=1}^n\delta_{\alpha_s,\alpha_{s+1}}+\frac{d}{D^n}.
	\lb{free7}
\en
Note that the first term in the right-hand side is larger but has the constraint on the $\alpha$'s while the second term is smaller but is (almost) free from constraint.
Going back to \rlb{TrBn1}, the second term yields a new contribution
\eq
	\sum_{\alpha_1, \ldots, \alpha_n=1}^D 
	A_{\alpha_1\alpha_2} \, A_{\alpha_2 \alpha_3} \, \cdots \,  
	A_{\alpha_n \alpha_1} \, 
	\frac{d}{D^n}
	=\tr[\sfA^n]\,\frac{d}{D^n},
	\lb{free8}
\en
to $\Av \bigl [ \tr [\sfB^n] \bigr ]$.
Since $\bigl|\tr[\sfA^n]\bigr|\le D\,\bmu^{n-1}$ by \rlb{Acond}, we find
\eq
\Bigl|\Av \bigl [ \tr [\sfB^n] \bigr ]\Bigr|\lesssim
D\,\Bigl(\frac{d}{D}\Bigr)^n+d\,\Bigl(\frac{\bmu}{D}\Bigr)^{n-1}
\lb{free9}
\en
where we again included the contribution \rlb{free5}.
This is basically the desired bound \rlb{trBn}, which is the main conclusion of Proposition~\ref{t:main3}.
In our application, where $\bmu/D\sim\tauB/\tau$, the second term in the right-hand side of \rlb{free9} becomes the main contribution\footnote{%
The first term becomes dominant when $\tau$ is extremely large.
} to $\Av \bigl [ \tr [\sfB^n] \bigr ]$.

%%%%%%%%%%%%%%%%%%%%%%%%%%
\subsubsection{Precise integration formula}
It is a nontrivial task to make the above heuristic estimate into a rigorous one.
In particular we have to treat the expectation value of $\xi$'s accurately, beyond the first order approximation \rlb{free2}.
Fortunately we can make use of the recent development due to Collins and others on the integration with respect to the Haar measure on the unitary group \cite{Coll03,CS06}.

We shall summarize the results which are relevant to us.
Let $d\mu_\mathrm{H}(\sfU)$ denote the Haar measure of the $D$-dimension unitary group $U(D)$, whose elements are matrices $\sfU=(U_\ab)_{\alpha,\beta\in\oD}$.
Then the integral of the matrix elements is given by Collins' formula
\eqa
\lb{UN-int.1}
	& 
	\int d\mu_\mathrm{H}(\sfU)\, \bigl ( U_{\alpha_1' \beta_1'} \, 
	U_{\alpha_2' \beta_2'} \, \cdots \, 
	U_{\alpha_n' \beta_n'}\,\bigr )^* \,\,
	U_{\alpha_1 \beta_1} \, U_{\alpha_2 \beta_2} \, 
	\cdots \, 
	U_{\alpha_n \beta_n}	
	\nl 
	& 
	= 
	\sum_{\sigma, \tau \in \Sn}
	I[\alpha_k = \alpha'_{\sigma(k)} \text{ and } \beta_{k} = 
	\beta'_{\tau(k)} \text{ for all } k= 1, 2, \ldots, n] \, 
	\Wg(\tau \sigma^{-1}),
\ena 
where $\Sn$ is the order-$n$ symmetric group, i.e., the set of all permutations of $\{1,2,\ldots,n\}$.  
See Theorem~2.2 of \cite{Coll03} and Corollary~2.4 of \cite{CS06}. 
Note that here (and in the following) $(U_\ab)^*$ means the complex conjugate of the complex number $U_\ab$, not the hermitian conjugate.

The Weingarten-Collins function $\Wg(\sigma)$ is a real valued function of $\sigma$, whose definition can be found in \cite{Coll03,CS06}. 
Its leading behavior for large $D$ is given by 
$|\Wg(\sigma)|\approx 1/D^{n+|\sigma|}$, where $|\sigma|$ denotes (throughout the present paper) the minimum number of transpositions necessary to express the permutation $\sigma$ as their products.
See Appendix~\ref{app-gp}.

We need an upper bound on $|\Wg(\sigma)|$ for our proof. 
We shall make use of the following, which behaves 
more or less similar to the above mentioned leading behavior.
%%%%% LEMMA %%%%%
\begin{lemma}
\label{lem-Weingarten-bd.1}
Let $n$ and $D$ satisfy $2 n^2 \le D$.
Then for any $\sigma \in \Sn$, we have
\eq
\lb{Weingartern-bd.1}
	|\Wg(\sigma)| \leq 
	\frac{2e\,n^{2|\sigma|}}{D^{n+|\sigma|}}
	=
	\frac{2e}{D^n}\Bigl(\frac{n^2}{D}\Bigr)^{|\sigma|}.
	\en
\end{lemma}
%%%%% LEMMA %%%%%

\bigno
{\em Proof:}\/
In the proof of Theorem~4 of \cite{CGGPG}, it is shown that
\eq
	\bigl |  \Wg(\sigma)  \bigr| \leq 
	\frac{n^{2|\sigma|}}{D^{n+|\sigma|}}e
	\sum_{k=0}^\infty\Bigl(\frac{n^2}{D}\Bigr)^k.
\en
We get \rlb{Weingartern-bd.1} by evaluating the summation.
\qedm

\bigskip

We note that the condition $2 n^2 \le D$ for Lemma~\ref{lem-Weingarten-bd.1} is always satisfied under the conditions of Proposition~\ref{t:main3}.
To see this note that $d\ge1$ implies $n^4\le2D$, which, along with $4\le n^2$, yields $2n^2\le D$.

Recall that our basis $\{\bsxi^{(j)}\}_{j=1,\ldots,d}$ is constructed from a random unitary transformation by $\bsxi^{(j)}=\hU\bspsi_j$.
If we define a unitary matrix $\sfU=(U_\ab)_{\alpha,\beta\in\oD}$ by $U_\ab:=\sbkt{\psa,\hU\psb}$, we have $U_{\alpha j}=\xi^{(j)}_\alpha$ where (as before) $\xi^{(j)}_\alpha$ is the coefficient in the expansion $\bsxi^{(j)}=\suma\xi^{(j)}_\alpha\psa$.
Then the expectation value in the summation in \rlb{aveP.11} is written as
\eqa
	\Av \bigl [\,&
	\xi^{(j_1)}_{\alpha_1} \,(\xi^{(j_1)}_{\alpha_2})^* \, 
	\xi^{(j_2)}_{\alpha_2} \,(\xi^{(j_2)}_{\alpha_3})^* \, 
	\cdots 
	\xi^{(j_n)}_{\alpha_{n}} \,
	(\xi^{(j_n)}_{\alpha_{1}})^* \, 
	\bigr ] 
	\nl&=
	\int d\mu_\mathrm{H}(\sfU)\, \bigl ( U_{\alpha_2 j_1} \, 
	U_{\alpha_3 j_2} \, \cdots \, 
	U_{\alpha_n j_{n-1}}\,U_{\alpha_{1}j_n}\bigr )^*  \,\,
	U_{\alpha_1 j_1} \, U_{\alpha_2 j_2} \, 
	\cdots \, 
	U_{\alpha_n j_n}.
\ena
The right-hand side is a special case of \rlb{UN-int.1}, where we have $\alpha'_k=\alpha_{k+1}$ and $\beta'_k=\beta_k=j_k$.
We thus find
\eq
	\Av \bigl [\,
	\xi^{(j_1)}_{\alpha_1} \,(\xi^{(j_1)}_{\alpha_2})^* \, 
	\cdots 
	\xi^{(j_n)}_{\alpha_{n}} \,
	(\xi^{(j_n)}_{\alpha_{1}})^* \, 
	\bigr ] 
	= \sum_{\sigma, \tau \in \Sn} 
	I[\forall k,  \alpha_{k} = \alpha_{\sigma(k)+1}\text{ and }
	j_k = j_{\tau(k)}  ] \, 
	\Wg(\tau \sigma^{-1}).
	\lb{Avxi=IW1}
\en
We wish to rewrite the constraint $\alpha_{k} = \alpha_{\sigma(k)+1}$ for any $k\in\{1,\ldots,n\}$ (where we always identify $n+1$ with 1) in a more convenient form.
Note that the condition is equivalent to $\alpha_{\sigma^{-1}(k')}=\alpha_{k'+1}$ for any $k'$, and hence to $\alpha_{\sigma^{-1}(\shif^{-1}(\ell))}=\alpha_\ell$ for any $\ell$, where $\shif$ is the shift defined by $\shif(\ell)=\ell+1$.
Thus by defining $\rho=\sigma^{-1}\shif^{-1}$, the condition can be rewritten as $\alpha_k=\alpha_{\rho(k)}$ for any $k$.
We can thus rewrite \rlb{Avxi=IW1} as
\eq
	\Av \bigl [\,
	\xi^{(j_1)}_{\alpha_1} \,(\xi^{(j_1)}_{\alpha_2})^* \, 
	\cdots 
	\xi^{(j_n)}_{\alpha_{n}} \,
	(\xi^{(j_n)}_{\alpha_{1}})^* \, 
	\bigr ] 
	= \sum_{\rho, \tau \in \Sn} 
	I[\forall k,  \alpha_{k} = \alpha_{\rho(k)}\text{ and }
	j_k = j_{\tau(k)}  ] \, 
	\Wg(\tau\rho\,\shif),
	\lb{Avxi=IW2}
\en
where we noted that $\sigma^{-1}=\rho\,\shif$

Substituting this back to \rlb{TrBn1} and \rlb{aveP.11}, we find that the quantity to be evaluated is rewritten and bounded as
\eqa
	\Bigl|\Av \bigl [ \tr [\sfB^n] \bigr ]\Bigr|
	& = \Biggl|\sum_{\alpha_1, \ldots, \alpha_n=1}^D
	\sum_{j_1, \ldots, j_n=1}^d
	\sum_{\rho, \tau \in \Sn} 
	A_{\alpha_1\alpha_2} \, A_{\alpha_2 \alpha_3} \, \cdots \,  
	A_{\alpha_n \alpha_1} \, 
	\nl&\hspace{1.5cm}
	\times I[\forall k,  \alpha_{k} = \alpha_{\rho(k)}\text{ and }
	j_k = j_{\tau(k)}  ] \, 
	\Wg(\tau\rho\,\shif)\Biggr|
	\nl
	&=
	\Biggl|\sum_{\rho\in\Sn}
	\Bigl\{\sum_{\alpha_1, \ldots, \alpha_n=1}^D
	A_{\alpha_1\alpha_2} \, A_{\alpha_2 \alpha_3} \, \cdots \,  
	A_{\alpha_n \alpha_1}
	I[\forall k,  \alpha_{k} = \alpha_{\rho(k)}]\Bigr\}
	\nl&\hspace{1cm}
	\times\Bigl\{
	\sum_{\tau\in\Sn}\Wg(\tau\rho\,\shif)
	\sum_{j_1, \ldots, j_n=1}^d
	I[\forall k, j_k = j_{\tau(k)}  ]
	\Bigr\}\Biggr|
	\nl
	&\le\sum_{\rho\in\Sn}|S_1(\rho)|\,|S_2(\rho)|,
	.
	\lb{TrBn2}
\ena
where we introduced two summations
\eq
	S_1(\rho):=\sum_{\alpha_1, \ldots, \alpha_n=1}^D
	A_{\alpha_1\alpha_2} \, A_{\alpha_2 \alpha_3} \, \cdots \,  
	A_{\alpha_n \alpha_1}
	I[\forall k,  \alpha_{k} = \alpha_{\rho(k)}],
	\lb{def-Sigma1}
\en
and
\eq
	S_2(\rho):=\sum_{\tau\in\Sn}\Wg(\tau\rho\,\shif)
	\sum_{j_1, \ldots, j_n=1}^d
	I[\forall k, j_k = j_{\tau(k)}  ].
	\lb{def-Sigma2}
\en
Now we shall separately bound $|S_1(\rho)|$ and $|S_2(\rho)|$.

%%%%%%%%%%%%%%%%%%%%%%%%%%
\subsubsection{Evaluation of $|S_1(\rho)|$}
\label{ss:EvS1}
For any permutation $\sigma\in\Sn$, let $i_\ell(\sigma)$ denote the number of distinct cycles of length $\ell$ in $\sigma$.
In particular $i_1(\sigma)$ denotes the number of $k\in\{1,\ldots,n\}$ such that $\sigma(k)=k$.
We also denote the total number of cycles in $\sigma$ as $c(\sigma):=\sum_{\ell=1}^ni_\ell(\sigma)$.
See Appendix~\ref{app-gp}.

Then we have the following bound for $|S_1(\rho)|$.
%%%%% LEMMA %%%%%
\begin{lemma}
\label{lemma-S1}
For any $\rho \in \Sn$, we have 
\eq
\lb{SumOverAlpha.95pre}
	|S_1 (\rho)|
	\leq 
	\begin{cases}
	D\,\bmu^{n-1}, & \text{\em if $\rho =\idp$};\\
	D^{c(\rho)} \, \Bigl(\dfrac{\bmu}{D}\Bigr )^{i_1(\rho)},
	& \text{\em if $\rho \neq \idp$}.
	\end{cases}
\en 
\end{lemma}
%%%%% LEMMA %%%%%

\bigskip
\noindent
{\em Proof:}\/
As a warm up, we consider two extreme cases.
First we set $\rho = \idp$, and observe that the constraint ``$\alpha_k = \alpha_{\rho(k)}$ for any $k$'' is satisfied for any choice of $\alpha_1,\ldots,\alpha_n$.
We then simply sum over $\alpha_1,\ldots,\alpha_n$ in \rlb{def-Sigma1} to get
\eq
\lb{SumOverAlpha.12}
	S_1(\idp)
	= \sum_{\alpha_1, \ldots, \alpha_n=1}^D
	A_{\alpha_1 \alpha_2} \, A_{\alpha_2 \alpha_3} 
	\cdots
	A_{\alpha_n \alpha_1}
	= \tr[\sfA^n].
\en
By using the assumption \rlb{Acond}, we get
\eq
	|S_1(\idp)|\le\suma\bigl|(\sfA^n)_{\alpha\alpha}\bigr|
	\le D\,\bmu^{n-1},
\en 
which proves \rlb{SumOverAlpha.95pre} for $\rho=\idp$.
It turns out that this is indeed the main contribution to the summation in the right-hand side of \rlb{TrBn2}.

The second extreme case is $\rho =\shif^{-1}$, which, in the original language in \rlb{Avxi=IW1}, corresponds to $\sigma=\idp$. 
Here the constraint reads ``$\alpha_k = \alpha_{k-1}$ for all $k$'', which in fact means 
$\alpha_1=\alpha_2=\cdots=\alpha_n$.
We thus find from \rlb{def-Sigma1} that
\eq 
\lb{SumOverAlpha.19}
	S_1(\shif^{-1}) = \sum_{\alpha=1}^D   
	A_{\alpha \alpha} \, A_{\alpha \alpha} 
	\cdots
	A_{\alpha \alpha},
\en
which, with the assumption  \rlb{Acond}, means
\eq
	|S_1(\shif^{-1})|\leq \sum_{\alpha=1}^D 1 = D.
\en
The bound \rlb{SumOverAlpha.95pre} is trivially satisfied because $c(\shif^{-1})=1$ and $i_1(\shif^{-1}) =0$.
We also note that this contribution corresponds to the crude (and indeed useless) estimate in \rlb{free1}--\rlb{free4}.

%%%%%FIGFIGFIGFIGFIGFIGFIGFIGFIGFIGFIGFIGFIGFIGFIGFIGFIGFIGFIGFIG
%%%%%FIGFIGFIGFIGFIGFIGFIGFIGFIGFIGFIGFIGFIGFIGFIGFIGFIGFIGFIGFIG
\begin{figure}[ht]
\begin{center}
\includegraphics[scale = 0.25]{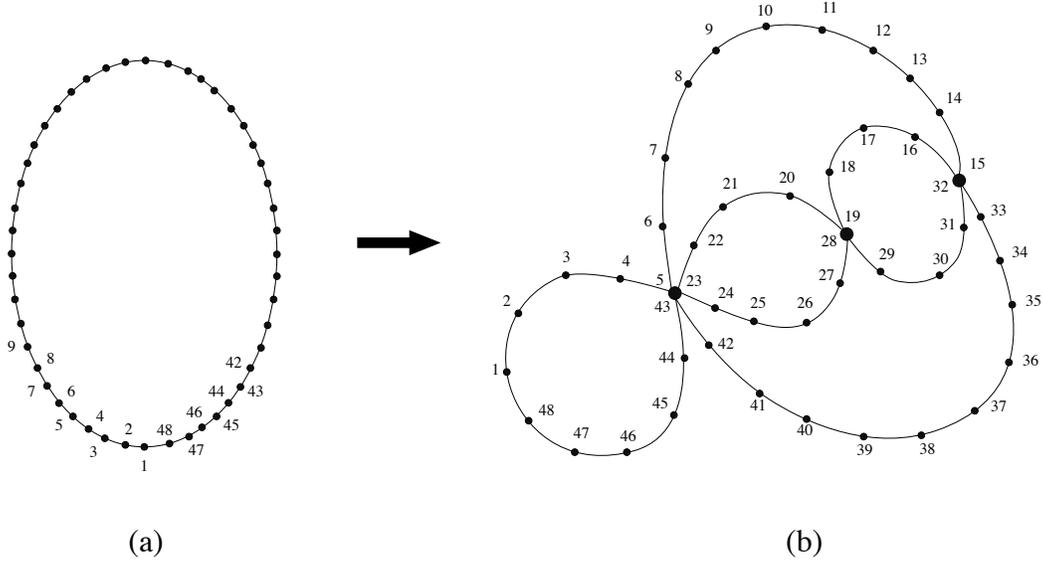} 
\end{center}
\caption{A diagram which represents the sum \rlb{def-Sigma1} for
$\rho = (5, 23, 43)(19,28)(15,32)$. (a) 
Without the constraint imposed by 
$I[\forall k,  \alpha_{k} = \alpha_{\rho(k)}]$, we have a simple single loop with 48 points on it. 
(b) With the constraint, we glue several points together to 
form the diagram on the right. Glued vertices 
($5=23=43, 19=28, 15=32$) are denoted by 
big dots. 
}
\label{fig-S1}
\end{figure}
%%%%%FIGFIGFIGFIGFIGFIGFIGFIGFIGFIGFIGFIGFIGFIGFIGFIGFIGFIGFIG
%%%%%FIGFIGFIGFIGFIGFIGFIGFIGFIGFIGFIGFIGFIGFIGFIGFIGFIGFIGFIG

To deal with a general permutation $\rho\ne\idp$, it is crucial to decompose $\rho$ into a product of disjoint cycles (see Appendix~\ref{app-gp}).
As an example let $n=48$ and consider the permutation $\rho = (5, 23, 43)(19,28)(15,32)$, where the three series of indices denote three cycles, and all the other indices are unchanged.
In this case, the constraint ``$\alpha_k = \alpha_{\rho(k)}$ for any $k$'' in \rlb{def-Sigma1} reads $\alpha_5 = \alpha_{23} = \alpha_{43}$, $\alpha_{19} = \alpha_{28}$, and $\alpha_{15} = \alpha_{32}$.

It is illuminating to represent the constraint diagrammatically as in Figure~\ref{fig-S1}.
We first represent
the product $A_{\alpha_1 \alpha_2} A_{\alpha_2 \alpha_3} 
\cdots A_{\alpha_{48} \alpha_{1}}$ as a 
simple loop with 48 points, where the points represent the indices $i=1,\ldots,48$, and the edges represent the elements
$A_{\alpha_i \alpha_{i+1}}$ with $i = 1,\ldots,48$ (where we identify 49 with 1) as in Figure~\ref{fig-S1}(a). 
To take into account the constraint, we identify (or glue together) those points corresponding to $\alpha$'s which are set equal.
Figure~\ref{fig-S1}(b) shows the resulting diagram.
A vertex with degree $\geq 4$ is 
called a {\em glued vertex}\/, and is denoted by a big dot. 
Those points which are not identified with others are called {\em single points}\/. 

Let us examine the summation \rlb{def-Sigma1}.
We first sum over all $\alpha$'s corresponding to single points.
Note that single points are located on lines whose end-points are glued vertices (or a glued vertex).
There are seven such lines in Figure~\ref{fig-S1}(b).
As an example, consider the line which contains $5$, $6$, $7,\ldots,14$, and  $15$.
Then the summation is readily evaluated as
\eq
	\sum_{\alpha_6, \alpha_7, \ldots, \alpha_{14}=1}^D 
	A_{\alpha_5 \alpha_6} A_{\alpha_6 \alpha_7} \cdots 
	A_{\alpha_{14} \alpha_{15}}
	=  ( \sfA^{10}  )_{\alpha_6 \alpha_{15}}
\en
which, again with the assumption  \rlb{Acond}, implies the upper bound
\eq
	\biggl|\sum_{\alpha_6, \alpha_7, \ldots, \alpha_{14}=1}^D 
	A_{\alpha_5 \alpha_6} A_{\alpha_6 \alpha_7} \cdots 
	A_{\alpha_{14} \alpha_{15}}\biggr|
	\leq (\bmu)^9. 
\en 
Note here that 9 is precisely the number of single points on the line.
This is true in general since the sum corresponding to a line with $\ell$ edges yields $(\sfA^\ell)_{\alpha\alpha'}$, which is bounded as $|(\sfA^\ell)_{\alpha\alpha'}|\le\bmu^{\ell-1}$ with $\ell-1$ being the number of single points on the line.

Recall that the number of single points on the whole diagram is equal to the number of fixed points of $\rho$, which is $i_1(\rho)$.
We thus find that the summation over all $\alpha$'s corresponding to single points gives
\eq
\Biggl|\biggl(\,\,\prod_{i:\text{single point}}\,\,\sum_{\alpha_i=1}^D\,\,\biggr)
A_{\alpha_1 \alpha_2} \, A_{\alpha_2 \alpha_3} 
\cdots A_{\alpha_n \alpha_1}\Biggr|
\le\bmu^{i_1(\rho)}.
\en

It remains to sum the upper bound $\bmu^{i_1(\rho)}$ over all $\alpha$'s on glued vertices.
We note that the number of the glued vertices is equal to the number of cycles with length greater than one, and is given by $c(\rho)-i_1(\rho)$.
We thus find
\eq
|S_1(\rho)|\le D^{c(\rho)-i_1(\rho)}\,\bmu^{i_1(\rho)}
=D^{c(\rho)}\Bigl(\frac{\bmu}{D}\Bigr)^{i_1(\rho)},
\en
which proves \rlb{SumOverAlpha.95pre} for $\rho\ne\idp$.
\qedm

%%%%%%%%%%%%%%%%%%%%%%%%%%
\subsubsection{Evaluation of $|S_2(\rho)|$}

%%%%% LEMMA %%%%%
\begin{lemma}
\label{lem-S2-bd.1}
Under the conditions  $n^4d\le 2D$ and $n\ge2$,
we have 
\eqa 
\lb{SumOverjs.49pre}
	|S_2(\rho)| 
	\leq 2 e^2 \,  D^{-n} \, d^{c(\rho\,\shif)}
	. 
\ena
\end{lemma}
%%%%% LEMMA %%%%%

\bigno
{\em Proof:}\/
For fixed $\tau$, the sum over $j_1,\ldots,j_n$ is evaluated to be
\eq
\lb{SumOverjs.17}
	\sum_{j_1,\ldots, j_n=1}^d \, 
	I[\forall k, \; j_k = j_{\tau(k)}]
	= d^{c(\tau)}.
\en
To see this we decompose $\tau$ into a product of disjoint cycles, and note that $j_k$ should be equal on each cycle in order to satisfy the constraint that $j_k = j_{\tau(k)}$ for any $k$.
Since the number of distinct cycles in $\tau$ is $c(\tau)$, we get \rlb{SumOverjs.17}.

By recalling the definition \rlb{def-Sigma2} of $S_2(\rho)$, and using the upper bound \rlb{Weingartern-bd.1} for the Weingarten-Collins function, we get
\eq
\lb{SumOverjs.19}
 	|S_2(\rho)|\le
	2e D^{-n} 
	\sum_{\tau\in \Sn} \Bigl ( \frac{n^2}{D} 
	\Bigr )^{|\tau\rho\,\shif|} \, 
	d^{c(\tau)}\, 
	. 
\en 
We define a new permutation $\omega:=\tau\rho\,\shif$, and sum over $\omega$ instead of $\tau$ to find
\eq
\lb{SumOverjs.33}
	|S_2(\rho)|\le2eD^{-n} 
	\sum_{\omega\in \Sn} \Bigl ( \frac{n^2}{D} 
	\Bigr )^{|\omega|} \, 
	d^{c(\omega\shif^{-1}\rho^{-1})}
	\le
	2eD^{-n}\,d^{c(\rho\,\shif)}
	\sum_{\omega\in \Sn} \Bigl ( \frac{n^2d}{D} 
	\Bigr )^{|\omega|},
\en
where we used
\eq
c(\omega\shif^{-1}\rho^{-1})\le|\omega|+c(\shif^{-1}\rho^{-1})
=|\omega|+c(\rho\,\shif),
\en
which is a consequence of \rlb{SumOverjs.35again}. 

To evaluate the sum over $\omega$, we let $a_{n,k}$ be the number of $\omega\in\Sn$ with $|\omega|=k$, and use the bound \rlb{ank-bd.1} where $N = n(n-1)/2$ to get
\eqa
	\sum_{\omega \in \Sn} \Bigl ( \frac{n^2d}{D} 
	\Bigr )^{|\omega|}
	&= \sum_{k=0}^{n-1} a_{n,k} \, 
	\Bigl ( \frac{n^2d}{D} 
	\Bigr )^{k}
	\leq 
	\sum_{k=0}^{n-1} \binom{N}{k} \Bigl ( \frac{n^2d}{D} 
	\Bigr )^{k}
	\nl&
	\leq 
	\sum_{k=0}^{n-1} \frac{N^k}{k!}\Bigl ( \frac{n^2 d}{D} 
	\Bigr )^{k}
	\leq \exp \Bigl ( \frac{n^2dN}{D} \Bigr ) 
	\leq e, 
\ena
where we noted that the condition $n^4d\le2D$ implies $n^2 d N \leq D$.
Substituting this bound into \rlb{SumOverjs.33}, we get the desired  \rlb{SumOverjs.49pre}.
\qedm

%%%%%%%%%%%%%%%%%%%%%%%%%%
\subsubsection{The sum over $\rho$}
We are ready to prove the desired bound \rlb{trBn} for $\Av[\tr[\sfB^n]]$.
We recall from \rlb{TrBn2} that
\eq
	\Bigl|\Av \bigl [ \tr [\sfB^n] \bigr ]\Bigr|
	\le\sum_{\rho\in\Sn}|S_1(\rho)|\,|S_2(\rho)|,
	\lb{TrBn3}
\en
and use the bounds obtained for the right-hand side.

\bigno
{\em Proof of Proposition~\ref{t:main3}:}\/
Substituting the bounds \rlb{SumOverAlpha.95pre} for $|S_1(\rho)|$ and \rlb{SumOverjs.49pre} for $|S_2(\rho)|$ into \rlb{TrBn3}, we find
\eqa 
\lb{SumOverTau.11}
	\Bigl|\Av \bigl [ \tr [\sfB^n] \bigr ]\Bigr|
	& \leq 
	2e^2d \Bigl ( \frac{\bmu}{D} \Bigr )^{n-1}
	+ 
	2e^2  \sumtwo{\rho \in \Sn}{(\rho \neq \idp)} D^{c(\rho)-n} \, 
	\Bigl ( \frac{\bmu}{D} \Bigr )^{i_1(\rho)} 
	\, d^{c(\rho\,\shif)}
	,
\ena
where we noted that $c(\shif)=1$.
By using \rlb{SumOverTau.23again}, which is
$c(\rho) + c(\rho\,\shif) \leq n+1$,
\rlb{SumOverTau.11} is bounded as
\eqa 
\lb{SumOverTau.31}
	\Bigl|\Av \bigl [ \tr [\sfB^n] \bigr ]\Bigr|
	& \leq 
	2 e^2d \Bigl ( \frac{\bmu}{D} \Bigr )^{n-1}
	+
	2e^2 \sumtwo{\rho \in \Sn}{(\rho \neq \idp)} D^{c(\rho)-n} \, 
	\Bigl ( \frac{\bmu}{D} \Bigr )^{i_1(\rho)}  
	\, d^{n +1 - c(\rho) }
	\nl
	& = 2 e^2d 
	\biggl [ 
	\Bigl ( \frac{\bmu}{D} \Bigr )^{n-1}
	+
	\sumtwo{\rho \in \Sn}{(\rho \neq \idp)} 
	\Bigl ( \frac{\bmu}{D} \Bigr )^{i_1(\rho)} \, 
	\Bigl ( \frac{d}{D} \Bigr )^{n - c(\rho) } \, 
	\biggr ].
\intertext{If we further use \rlb{|rho|andc(rho)}, 
which is $c(\rho) = n - |\rho|$, the upper bound becomes} 
	& 
	= 2 e^2d 
	\biggl [ 
	\Bigl ( \frac{\bmu}{D} \Bigr )^{n-1}
	+
	\sumtwo{\rho \in \Sn}{(\rho \neq \idp)} 
	\Bigl ( \frac{\bmu}{D} \Bigr )^{i_1(\rho)} \, 
	\Bigl ( \frac{d}{D} \Bigr )^{|\rho|} \, 
	\biggr ] \, 
	.
\ena
We shall bound the sum over $\rho$ in the right-hand side.
In the following, the inequality \rlb{|rho|and-i1}, which is
\eq	
\lb{SumOverTau.51againagain}	
	2|\rho| + i_1(\rho) \geq n,
\en
plays an important role.
To simplify notation let us write $\alphan:=\bmu/D$ and $\betan:=d/D$, and note that the assumption $n^{2+(4/\delta)}d\le D$ (in the statement of Proposition~\ref{t:main3}) implies  $N^{1+2/\delta}\betan\le1$, where $N=n(n-1)/2$.
We treat the following three cases separately.

(i) $\betan \leq \alphan^{2+\delta}$ and $\alphan\le1$:
We use the trivial identity 
$\betan = \betan^{2/(2+\delta)}\times\betan^{\delta/(2+\delta)}$ 
together with 
\rlb{SumOverTau.51againagain} to get 
\eqa 
	\alphan^{i_1(\rho)} \betan^{|\rho|} 
	& = \alphan^{i_1(\rho)} \bigl ( 
	\betan^{2/(2+\delta)} \bigr)^{|\rho|} 
	\bigl( \betan^{\delta/(2+\delta)}\bigr )^{|\rho|}
	\leq 
	\alphan^{i_1(\rho)} \alphan^{2|\rho|} 
	\bigl( \betan^{\delta/(2+\delta)}\bigr )^{|\rho|}
	\nl
	& 
	= 
	\alphan^{i_1(\rho) + 2|\rho|} 
	\bigl( \betan^{\delta/(2+\delta)}\bigr )^{|\rho|}
	\leq \alphan^{n}
	\bigl( \betan^{\delta/(2+\delta)}\bigr )^{|\rho|}
	. 
\ena 
We then use \rlb{ank-bd.1} to bound the number of $\rho$
with $|\rho|=k$ (note that $\rho \neq \idp$ means $k \geq 1$) to find 
\eqa 
	\sumtwo{\rho \in \Sn}{(\rho \neq \idp)} 
	\alphan^{i_1(\rho)} \, \betan^{|\rho|}
	& \leq 
	\alphan^{n} 
	\sumtwo{\rho \in \Sn}{(\rho \neq \idp)} 
	\bigl( \betan^{\delta/(2+\delta)}\bigr )^{|\rho|}
	\leq 
	\alphan^{n} \, 
	\sum_{k=1}^{n-1} \binom{N}{k} 
	\bigl( \betan^{\delta/(2+\delta)}\bigr )^{k}
	\leq \alphan^{n} \, 
	\sum_{k=1}^{n-1}  
	\frac{N^k}{k!}
	\bigl( \betan^{\delta/(2+\delta)}\bigr )^{k}
	\nl 
		& 
	=
	\alphan^{n} \, 
	\sum_{k=1}^{n-1}  
	\frac{1}{k!}
	\bigl(N\, \betan^{\delta/(2+\delta)}\bigr )^{k}
	\leq 
	\alphan^{n} \, \Bigl \{ \exp \bigl ( 
	N \, \betan^{\delta/(2+\delta)} \bigr ) -1 \Bigr \} 
	\leq (e-1) \, \alphan^{n}, 
	\lb{SumOverTau.63}
\ena
where we used $N \, \betan^{\delta/(2+\delta)} \leq 1$ 
or, equivalently, $N^{1+2/\delta} \, \betan \leq 1$. 

(ii) $\betan \leq \alphan^{2+\delta}$ and $\alphan\ge1$:
This case is trivial.
Noting that $i_1(\rho)\le n$, we have
\eqa 
	\sumtwo{\rho \in \Sn}{(\rho \neq \idp)} 
	\alphan^{i_1(\rho)} \, \betan^{|\rho|}
	& \leq  
	\alphan^n\sum_{k=1}^{n-1} \binom{N}{k}\betan^k
	\leq
	\alphan^n\sum_{k=1}^{n-1} \frac{(N\betan)^k}{k!}
	\leq (e-1) \, \alphan^{n}, 
	\lb{SumOverTau.63B}
\ena
because $N\betan\leq N^{1+2/\delta} \, \betan \leq 1$.

(iii) $\betan \geq \alphan^{2+\delta}$:
Here we simply bound $\alphan$ as $\alphan\le\betan^{1/(2+\delta)}$, and 
use \rlb{SumOverTau.51againagain} to get 
\eqa 
	\alphan^{i_1(\rho)} \, \betan^{|\rho|}
	& \leq 
	\bigl (\betan^{1/(2+\delta)} \bigr)^{i_1(\rho)} \, \betan^{|\rho|}
	= 
	\bigl (\betan^{1/(2+\delta)} \bigr)^{i_1(\rho)+2|\rho|+\delta|\rho|}
	\leq 
	(\betan^{1/(2+\delta)} \bigr)^{n} 
	(\betan^{\delta/(2+\delta)} \bigr)^{|\rho|}
	. 
\ena
Since the sum of $(\betan^{\delta/(2+\delta)} \bigr)^{|\rho|}$ over $\rho\in\Sn\backslash\{\idp\}$ has been shown to be less than or equal to $e-1$ in \rlb{SumOverTau.63}, we find
\eq
\lb{SumOverTau.75}
	\sumtwo{\rho \in \Sn}{(\rho \neq \idp)} 
	\alphan^{i_1(\rho)} \, \betan^{|\rho|}
	\leq (e-1) \, (\betan^{1/(2+\delta)} \bigr)^{n}.
\en

Combining \rlb{SumOverTau.63}, \rlb{SumOverTau.63B} and \rlb{SumOverTau.75}, we finally get
\eq
\lb{SumOverTau.71}
	\sumtwo{\rho \in \Sn}{(\rho \neq \idp)} 
	\alphan^{i_1(\rho)} \, \betan^{|\rho|}
	\leq 
	(e-1) \,  \Bigl ( \alphan \vee \betan^{1/(2+\delta)} \Bigr )^{n}
	=  (e-1) \, \biggl \{ \frac{\bmu}{D}  \vee 
	\Bigl ( \frac{d}{D} \Bigr )^{1/(2+\delta)}\biggr \}^{n}
	. 
\en 
Going back to \rlb{SumOverTau.31}, we obtain
\eq
	\Bigl|\Av \bigl [ \tr [\sfB^n] \bigr ]\Bigr|
	\leq 
	2e^2d \, 
	\biggl [ 
	\Bigl ( \frac{\bmu}{D} \Bigr )^{n-1} 
	+ (e-1) \, \biggl \{ \frac{\bmu}{D} \vee 
	\Bigl ( \frac{d}{D} \Bigr )^{1/(2+\delta)}\biggr \}^{n}\, 
	\biggr ] 
	,
\en 
which is the main inequality \rlb{trBn} in Proposition~\ref{t:main3}.
\qedm

\appendix
%%%%%%%%%%%%%%%%%%%%%%%%%%%%%%%%%%%%%%%%%%%%%%%%%%%%%%%%%%%%%%%%%%%%%%
%%%%%%%%%%%%%%%%%%%%%%%%%%%%%%%%%%%%%%%%%%%%%%%%%%%%%%%%%%%%%%%%%%%%%%
%%%%%%%%%%%%%%%%%%%%%%%%%%%%%%%%%%%%%%%%%%%%%%%%%%%%%%%%%%%%%%%%%%%%%%
%%%%%%%%%%%%%%%%%%%%%%%%%%%%%%%%%%%%%%%%%%%%%%%%%%%%%%%%%%%%%%%%%%%%%%
%%%%%%%%%%%%%%%%%%%%%%%%%%%%%%%%%%%%%%%%%%%%%%%%%%%%%%%%%%%%%%%%%%%%%%
%%%%%%%%%%%%%%%%%%%%%%%%%%%%%%%%%%%%%%%%%%%%%%%%%%%%%%%%%%%%%%%%%%%%%%

%%%%%%%%%%%%%%%%%%%%%%%%%%%%%%%%%%%%%%%%%%%%%%%%%%%%%%%%%%%%%%%%%%%%%%
%%%%%%%%%%%%%%%%%%%%%%%%%%%%%%%%%%%%%%%%%%%%%%%%%%%%%%%%%%%%%%%%%%%%%%
\section{Two theorems on thermalization}
\label{app-thermalization}
We shall state two theorems which show that general quantum systems thermalize under suitable assumptions.
As we have noted in the footnote~\ref{fn:twoclasses} (page~\pageref{fn:twoclasses}), such results can be roughly divided into two classes, which are essentially different.
The two theorems may be regarded as representatives of these classes\footnote{%
Like the main body of the present paper, these theorems use the notion of equilibrium based on the decomposition of the Hilbert space.
This formulation seems to be essential for the theorems.
As we noted in the footnote~\ref{fn:thermalizationEquilibration}, most of the existing works use different formulations for thermalization or equilibration, to which the following arguments do not apply.
}.

Here we take the same setting as in section~\ref{s:BG}.
The nonequilibrium subspace $\Hneq$ is a fixed $d$-dimensional subspace of the $D$-dimensional microcanonical energy shell $\calH$.
We further assume that all the energy eigenvalues (within the microcanonical energy shell) are nondegenerate, i.e., $\alpha\ne\beta$ implies $E_\alpha\ne E_\beta$ for any $\alpha,\beta\in\oD$.

The first theorem relies on the assumption of the  ``energy eigenstate thermalization''.
It was essentially first proved by von Neumann \cite{vonNeumann} (see also \cite{GLMTZ10,GLTZ}) as an easy part of his deep results.
%%%%% THM %%%%%
\begin{theorem}
\label{t:app-th-1}
Assume  that there is a (small) constant $\varepsilon>0$ such that one has
\eq
	\sbkt{\psa,\Pneq\,\psa}\le\varepsilon,
	\lb{EET1}
\en
for any $\alpha\in\oD$.
Then for any initial state $\varphi(0)\in\calH$,
\eq
	\lim_{\tau\up\infty}
	\frac{1}{\tau}\int_0^\tau dt\,\sbkt{\bsphi(t),\Pneq\,\bsphi(t)}
	\le
	\varepsilon.
	\lb{app-th-1}
\en
\end{theorem}
%%%%% THM %%%%%

A notable point of the theorem is that one is allowed to take any initial state from the energy shell.
Similar results are obtained in \cite{GLMTZ09b,Hal2010} and in the present work.

The second theorem, which was first stated in our unpublished work \cite{GHT2ndLaw}, is interesting in the sense that we do not need any assumptions like the ``energy eigenstate thermalization''.
Instead, we make an assumption on the initial state, one that can readily be satisfied when the dimension $d$ of $\Hneq$ is much smaller than the dimension $D$ of $\calH$.
%%%%% THM %%%%%
\begin{theorem}
\label{t:app-th-2}
Take an arbitrary initial state $\varphi(0)\in\calH$ which satisfies
\eq
	\bigl|\sbkt{\psa,\varphi(0)}\bigr|^2\le\frac{\varepsilon}{d},
	\lb{app-th-2-cond}
\en
for any $\alpha\in\oD$, where $\varepsilon>0$ is a (small) constant.
Then
\eq
	\lim_{\tau\up\infty}
	\frac{1}{\tau}\int_0^\tau dt\,\sbkt{\bsphi(t),\Pneq\,\bsphi(t)}
	\le
	\varepsilon.
	\lb{app-th-2}
\en
\end{theorem}
%%%%% THM %%%%%

The assumption \rlb{app-th-2-cond} basically means that the initial state $\bsphi(0)$ is distributed over many different energy eigenstates.
Since $D\gg d$, one can take small $\varepsilon$ such that\footnote{
One can, for example, set $\varepsilon=\sqrt{d/D}$.
} $\varepsilon/d\gg1/D$, which means that there are plenty of $\bsphi(0)$ satisfying the assumption.

Such an approach to thermalization using the initial state with a broad energy distribution starts, as far as we know, from \cite{Hal1998} and includes many works such as \cite{Reimann,LindenPopescuShortWinter,ReimannKastner,Reimann2}.

We also note that \rlb{app-th-1} or \rlb{app-th-2} readily implies
\eq
	\frac{1}{\tau}\int_0^\tau dt\,\sbkt{\bsphi(t),\Pneq\,\bsphi(t)}
	\le
	2\varepsilon,
\en
for a sufficiently large $\tau>0$, where how large $\tau$ should be depends on the initial state $\bsphi(0)$.
We have thus proved \rlb{approach}.

\bigno
{\em Proof of Theorems~\ref{t:app-th-1} and \ref{t:app-th-2}:}\/ We expand the initial state $\bsphi(0)$ as in \rlb{phi0exp}.
Then one easily finds
\eq
\frac{1}{\tau}\int_0^\tau dt\,\sbkt{\bsphi(t),\Pneq\,\bsphi(t)}
=\sumab(c_\alpha)^*c_\beta\,\frac{1}{\tau}\int_0^\tau dt\,
e^{i(E_\alpha-E_\beta)t/\hbar}
\sbkt{\psa,\Pneq\,\psb}.
\lb{app-th-AV}
\en
Since the energy eigenvalues are assumed to be nondegenerate, the $\tau\up\infty$ limit becomes
\eq
\lim_{\tau\up\infty}\frac{1}{\tau}\int_0^\tau dt\,\sbkt{\bsphi(t),\Pneq\,\bsphi(t)}
=\suma|c_\alpha|^2\,\sbkt{\psa,\Pneq\,\psa}.
\lb{app-th-AVti}
\en
We shall show that, under the assumptions of each theorem, the right-hand side does not exceed $\varepsilon$.

Under the assumption of Theorem~\ref{t:app-th-1}, the right-hand side of \rlb{app-th-AVti} is readily bounded as
\eq
\suma|c_\alpha|^2\,\sbkt{\psa,\Pneq\,\psa}\le
\suma|c_\alpha|^2\,\varepsilon=\varepsilon.
\en
Under the assumption of Theorem~\ref{t:app-th-2}, it is bounded as
\eq
\suma|c_\alpha|^2\,\sbkt{\psa,\Pneq\,\psa}\le
\frac{\varepsilon}{d}\suma\sbkt{\psa,\Pneq\,\psa}
=\frac{\varepsilon}{d}\,{\rm Tr}[\Pneq]=\varepsilon,
\en
where we noted that $c_\alpha=\sbkt{\psa,\varphi(0)}$ and ${\rm Tr}[\Pneq]=d$.
\qedm

%%%%%%%%%%%%%%%%%%%%%%%%%%%%%%%%%%%%%%%%%%%
%%%%%%%%%%%%%%%%%%%%%%%%%%%%%%%%%%%%%%%%%%%
\section{On the density of states}
\label{app-density}

Here we shall explain a way to explicitly construct the density of states.
We also justify the formula \rlb{sumtoint}, which allows one to convert a summation into an integral.

Suppose that the energy eigenvalues $E_1,\ldots,E_D\in[U-\DU,U]$ are given.
We define a smooth function\footnote{%
	Any normalized smooth function with a finite support centered at the 
	origin can be used as $k(x)$.
	We here stick to this choice just for concreteness.
}% 
\eq
	k(x) := \begin{cases} \calN\, \exp \{- (1-x^2)^{-1} \},
	& \text{ if } \quad |x| < 1; \\
	0, & \text{ otherwise},
	\end{cases}
\en 
where $\calN\sim2.2523$ is a normalization factor introduced 
to realize $\int k(x) dx = 1$. 
For any $\eta>0$, we define
\eq
	\rho_{\eta}(E) := \sum_{\alpha=1}^{D} \frac{1}{\eta} \, 
	k \Bigl ( \frac{E-E_\alpha}{\eta} \Bigr ). 
\en
Then we can easily show (see below) for any differentiable function $f(E)$ that
\eq
	\frac{1}{D}\biggl|\suma f(E_\alpha)-\int_{U-\DU-\eta}^{U+\eta} dE\,
	\rho_\eta(E)\,f(E)\biggr|
	\le \eta \times \sup_{E\in[U-\DU-\eta,U+\eta]}|f'(E)|,
\lb{sumtoint2}
\en
which is \rlb{sumtoint} when $\rho_\eta(E)$ is $\rho(E)$.

By definition, $\rho_\eta(E)$ is a summation of many functions, each 
having a width of order $\eta$ and a height of order $1/\eta$.  
Note that the limit $\rho_0(E)$ is a collection of $\delta$ functions, and 
exactly describes the discrete distribution of $E_\alpha$'s. 
The idea is to use an appropriate $\eta >0$ and get 
$\rho_\eta(E)$ that can be regarded as a physical density of states, which is smooth and monotone increasing.

If $\eta$ is too small, $\rho_\eta(E)$ still looks like a collection 
of $\delta$ functions, and is far from monotone.
It does not look like a usual density of states, and we can never hope for the bound \rlb{rEbD} to be valid.
If $\eta$ is sufficiently larger than the typical 
spacing of the energy levels, on the other hand, we expect 
a monotone 
$\rho_\eta$ which behaves like a density of states, 
and for which \rlb{rEbD} is justified.  
Because we have $D$ levels, we expect to have 
a well-behaved $\rho_\eta(E)$ for $\eta = D^{-\kappa}\tbeta^{-1}$ with $\kappa <1$ provided that $D \gg 1$.
The constant $\tbeta$ is introduced so that $\eta$ has the dimension of energy.

\bigno
{\em Proof of}\/ \rlb{sumtoint}:
Because $k(x)$ is normalized, we can write 
\eqalign 
	\sum_{\alpha=1}^{D} f(E_\alpha) 
	= \sum_{\alpha=1}^{D} \int dE \, \frac{1}{\eta} \, k \Bigl ( 
	\frac{E-E_\alpha}{\eta} \Bigr ) \, f(E_\alpha) 
	. 
\enalign  
Therefore, the left hand side of \rlb{sumtoint}, multiplied by $D$, 
can be written as 
\eqalign 
	\sum_{\alpha=1}^{D} f(E_\alpha) - \int dE \, \rho_\eta(E) \, f(E) 
	= \sum_{\alpha=1}^{D} \int dE \, \frac{1}{\eta} \, k \Bigl ( 
	\frac{E-E_\alpha}{\eta} \Bigr ) \, 
	\, \{f(E_\alpha) - f(E) \} 
	.
\enalign 
Now by the mean value theorem, 
\eq
	\bigl | \, f(E_\alpha) - f(E) \, \bigr | 
	\leq |f'(E^*) | \, \, |E_\alpha - E| 
	\leq \bar{f} \, |E_\alpha - E| , 
\en 
where $E^*$ is a point between $E_\alpha$ and $E$, and we 
introduced 
$\displaystyle \bar{f} := \sup_{U-\DU-\eta \leq E \leq U+\eta} |f'(E)|$. 
We thus have 
\eqalign 
	& 
	\Bigl | \, 
	\sum_{\alpha=1}^{D} f(E_\alpha) - \int dE \, \rho_\eta(E) \, f(E)
	\Bigr | 
	\leq 
	\sum_{\alpha=1}^{D} \int dE \, \frac{1}{\eta} \, k \Bigl ( 
	\frac{E-E_\alpha}{\eta} \Bigr ) \, 
	\, \bar{f} \, |E_\alpha - E|  
	\nnb 
	& = D \bar{f} \, \int dE \, \frac{1}{\eta} \, k \Bigl ( 
	\frac{E-E_\alpha}{\eta} \Bigr ) \, |E_\alpha - E|
	= D \bar{f} \eta \int dy \, k(y) \, |y| 
	\leq D \bar{f} \eta , 
\enalign 
where in the last step we used the fact that the expectation of 
$|y|$ with respect to the normalized 
measure $d y \, k(y)$ cannot exceed one. 
\qedm

%%%%%%%%%%%%%%%%%%%%%%%%%%%%%%%%%%%%%%%%%%%%%%%%%%%%%%%%%%%%%%%%%%%%%%
%%%%%%%%%%%%%%%%%%%%%%%%%%%%%%%%%%%%%%%%%%%%%%%%%%%%%%%%%%%%%%%%%%%%%%
\section{Some elementary facts about the symmetric group}
\label{app-gp}
We shall here summarize some elementary facts about the symmetric group that we used in the proof.
Although most (or all) of these facts may be well-known to experts, we include the proofs of some statements for completeness.

By $\Sn$ we denote the symmetric group of order $n$, i.e., a group consisting of all permutations of $\{1, \ldots, n\}$.

Any $\sigma\in\Sn$ can be written as a product of transpositions.
By $|\sigma|$ we denote the minimum number of transpositions needed in such a representation.

A {\em cycle}\/ (or a {\em cyclic permutation}\/) of length $\ell$ is a special permutation in which only $\ell$ indices $j_1,\ldots,j_\ell\in\{1,\ldots,n\}$ are changed in a cyclic manner, i.e., $j_k\to j_{k+1}$ (where we identify $j_{\ell+1}$ with $j_1$).
It is abbreviated as $(j_1,\ldots,j_\ell)$.

Any $\sigma\in\Sn$ can be decomposed into a product of disjoint cycles.
The decomposition is unique up to the ordering of the cycles.
We denote by $i_\ell(\sigma)$ the number of cycles of length $\ell$ in this decomposition.
We define $i_1(\sigma)$ to be the number of indices $j$ such that $\sigma(j)=j$.
Note the trivial identity $n=\sum_{\ell=1}^n\ell\,i_\ell(\sigma)$.
We also denote by 
\eq
\lb{SymGp.11}
	c(\sigma) = \sum_{l=1}^{n} i_\ell (\sigma)
\en
the total number of cycles in $\sigma$.

Let $\ell\ge2$ and $\tau$ be a cycle of length $\ell$.
One finds by inspection that $|\tau|=\ell-1$.
This fact implies for general $\sigma\in\Sn$ the well known identity
\eq	
\lb{|rho|andc(rho)}
	|\sigma| + c(\sigma) = n. 
\en

%%%%% Lemma %%%%%
\begin{lemma}
\label{lem-|tau|-and-i_1(tau)}
For any $\sigma \in \Sn$, one has
\eq	
\lb{|rho|and-i1}
	2 |\sigma| + i_1(\sigma) \geq n 
	.
\en 
\end{lemma}
%%%%% Lemma %%%%%
\bigno
{\em Proof:}\/ Simply note that
\eqa 
	|\sigma| 
	& 
	= n-c(\sigma) = \sum_{l=1}^{n} \ell \, i_\ell(\sigma) 
	- \sum_{\ell=1}^{n} i_\ell(\sigma)
	= \sum_{\ell=2}^{n} (\ell-1)\,i_\ell(\sigma)
	\nl&
	\geq \sum_{\ell=2}^{n} \frac{\ell}{2} \, i_\ell(\sigma)
	= \frac{1}{2} \biggl\{
	\Bigl(\sum_{\ell=1}^{n} \ell \, i_\ell(\sigma)\Bigr) - i_1(\sigma) 
	\biggr\}
	= \frac{1}{2} \{n - i_1(\sigma)\}.
\ena
\qedm

%%%%% LEMMA %%%%% 
\begin{lemma}
\label{lem-ochiai1}
For any permutations $\sigma$ and $\rho$, one has
\eq
\lb{SumOverTau.21again}
	c(\sigma) + c(\rho) \leq n + c(\sigma\rho)
	. 
\en 
\end{lemma}
%%%%% LEMMA %%%%%
If $\shif$ denotes the shift defined by $\shif(k)=k+1$ (where we identify $n+1$ with 1), then  \rlb{SumOverTau.21again} implies for any $\rho\in\Sn$ that
\eq
\lb{SumOverTau.23again}
	c(\rho) + c(\rho\,\shif)=c(\rho\,\shif)+c(\rho^{-1}) \leq n+c(\shif)=n+1
	. 
\en 

\bigno
{\em Proof of Lemma~\ref{lem-ochiai1}:}\/ 
By \rlb{|rho|andc(rho)}, we have
\eq
	c(\sigma) + c(\rho) = 2n - |\sigma| - |\rho|. 
\en 
But we trivially have
\eq
 |\sigma\rho |\le	|\sigma| + |\rho|,
\en 
since we can construct $\sigma\rho$, by using transpositions 
used for $\sigma$ and $\rho$.
Therefore, 
\eq
	c(\sigma) + c(\rho) = 2n - |\sigma| - |\rho|
	\leq 2n - |\sigma\rho | = n + c(\sigma\rho),
\en 
where we again used \rlb{|rho|andc(rho)}.
\qedm

%%%%% LEMMA %%%%%  
\begin{lemma}
\label{lem-rho-and-omega}
For any $\omega,\sigma\in\Sn$, one has
\eq
\lb{SumOverjs.35again}
	c(\omega \sigma) \leq 
	|\omega|+c(\sigma).
\en 
In particular, if $\tau$ is a transposition, 
\eq
\lb{SumOverjs.36again}
	c(\tau \sigma) \leq 
	c(\sigma) + 1 .  
\en 
\end{lemma}
%%%%% LEMMA %%%%%

\bigno
{\em Proof:}\/
We shall prove \rlb{SumOverjs.36again}.
Then \rlb{SumOverjs.35again} follows by decomposing $\omega$ into $|\omega|$ transpositions and by using \rlb{SumOverjs.36again} repeatedly.

We decompose $\sigma$ into disjoint cycles as 
$\sigma = c_1 c_2 \ldots c_m$ where $m=c(\sigma)$, 
and examine the decomposition of $\tau\sigma$ into disjoint cycles.
There are two cases.
(i)~Suppose that $\tau$ interchanges two elements in a single cycle $c_i$.
Then one easily finds (by explicit construction) that $c_i$ splits into two disjoint cycles by the action of $\tau$.
We thus have $c(\tau\sigma)=c(\sigma)+1$.
(ii)~Suppose that $\tau$ interchanges an element in $c_i$ and an element in $c_j$, where $i\ne j$.
Again one can easily check that the action of $\tau$ merges $c_i$ and $c_j$ to form a bigger single cycle.
We thus have $c(\tau\sigma)=c(\sigma)-1$.
\qedm

%%%%% LEMMA %%%%%  
\begin{lemma}
\label{lem-of-sigma-|sigma|}
Let $a_{n,k}$ be the number of $\sigma\in \Sn$ with $|\sigma|=k$.  
Then we have
\eq
\lb{ank-bd.1}
	a_{n,k} \leq \binom{N}{k},
\en 
where $N:=n(n-1)/2$.
\end{lemma}
%%%%% LEMMA %%%%%

\bigno
{\em Proof:}\/ 
Any permutation $\sigma$ with $|\sigma|=k$ is a
product of $k$ transpositions. 
By fixing a certain canonical ordering (that takes into account the cyclic structure of permutations), the order of the transpositions in such a product can in fact be determined merely by the set of $k$ transpositions involved in the product.
Since the number of possible transpositions is $N$, the number $a_{n,k}$ does not exceed $\binom{N}{k}$, which is the number of all possible choices of $k$ transpositions.
\qedm

\bigskip
\bigskip

We wish to thank Hiroyuki Ochiai for his indispensable 
help in the issues regarding the symmetric group, such as 
Lemmas~\ref{lem-ochiai1} and \ref{lem-of-sigma-|sigma|}, and 
 Beno\^{i}t Collins for discussions about 
upper bounds on $|\Wg(\sigma)|$, and for informing us of 
the literature \cite{CGGPG}. 
We also thank
Tetsuo Deguchi,
Takaaki Monnai,
Shin-ichi Sasa,
Akira Shimizu,
Tomoyuki Shirai,
Ayumu Sugita, 
Sho Sugiura,
and
Yu Watanabe
for valuable discussions and comments,
and Marcus Cramer for bringing our attention to \cite{Cramer}.

The present work was  supported in part by grant no.~37433 from the John Templeton Foundation (S.G.), JSPS Grants-in-Aid for Scientific Research nos.~25610021 (T.H.) and 25400407 (H.T.).

%%%%%%%%%%%%%%%%%%%%%%%%%%%%%%%%%%%%%%%%%%%
%%%%%%%%%%%%%%%%%%%%%%%%%%%%%%%%%%%%%%%%%%%

%%%%%%%%%%%%%%%%%%%%%%%%%%%%%%%%%%%%%%
%%%%%%%%%%%%%%%%%%%%%%%%%%%%%%%%%%%%%%
%%%%%%%%%%%%%%%%%%%%%%%%%%%%%%%%%%%%%%
\end{document}